\documentstyle [11pt]{article}
\newcounter{zaehler} \newcounter{notes} \newcounter{note}

\begin{document}

\title{\large\bf CELLULAR AUTOMATON MODEL OF \\ REACTION-TRANSPORT PROCESSES
\vspace{5mm} }
\author{T. Karapiperis, \\
{\em Paul Scherrer Institute, CH-5232 Villigen PSI, Switzerland} \vspace{7mm}
\\ B. Blankleider, \\
{\em Paul Scherrer Institute, CH-5232 Villigen PSI, Switzerland,} \\
{\em and School of Physical Sciences, Flinders University, Bedford Park,
S.A. 5042, Australia}\thanks{Present address} }

\date{ }
\maketitle

\begin{abstract}
\setlength\baselineskip{24pt}
\indent

The transport and chemical reactions of solutes are modelled as a cellular
automaton in which molecules of different species perform a random walk on a
regular lattice and react according to a local probabilistic rule. The model
describes advection and diffusion in a simple way, and as no restriction is
placed on the number of particles at a lattice site, it is also able to
describe a wide variety of chemical reactions. Assuming molecular chaos and a
smooth density function, we obtain the standard reaction-transport equations in
the continuum limit. Simulations on one- and two-dimensional lattices show that
the discrete model can be used to approximate the solutions of the continuum
equations. We discuss discrepancies which arise from correlations between
molecules and how these disappear as the continuum limit is approached. Of
particular interest are simulations displaying long-time behaviour which
depends
on long-wavelength statistical fluctuations not accounted for by the standard
equations. The model is applied to the reactions $a + b \rightleftharpoons c$
and $a + b \rightarrow c$ with homogeneous and inhomogeneous initial conditions
as well as to systems subject to autocatalytic reactions and displaying
spontaneous formation of spatial concentration patterns.

\end{abstract}

\setlength\baselineskip{24pt}

\newpage
\section{Introduction}   \label{intro}
\indent

The transport of aqueous solutions of contaminants in geological media is
inextricably coupled with a rich variety of physical and chemical processes.
The solutes may decay radioactively, react chemically with each other, sorb on
solid surfaces or change the porosity of the host rock by
precipitation/dissolution. Modelling an aqueous system in its full complexity
constitutes, conceptually and mathematically, a formidable task. A model is
essentially defined by selecting, on grounds of usefulness and economy, an
appropriate set of dependent variables (e.g.\ aqueous species concentrations)
and writing down the laws (e.g.\ mass conservation, law of mass action) these
variables satisfy. A model intended for practical applications has to be
translated subsequently into a usable and efficient computer code.

The model presented here takes advantage of the fact that the migration and
chemical transformation of aqueous species consists, at the microscopic level,
of  processes taking place {\em in parallel} (i.e.\ simultaneously at many
locations) and {\em locally} (i.e.\ involving molecules within a small spatial
neighbourhood). Each species is represented by a (large) number of `particles'
that move and react according to simple rules mimicking the microscopic
behaviour
of the  actual molecules. The model is implemented as simple computer code
suitable for  {\em massively parallel computers}. Our approach contrasts with
the
standard  modelling approach which solves systems of non-linear {\em partial
differential  equations (PDE's)} for macroscopic variables. The latter approach
neglects  microscopic details by effectively averaging solute properties over a
small macroscopic volume and dealing only with their local mean values.

Reaction-transport processes are modelled here, following Refs.~\cite{b,kb}, as
a {\em cellular automaton (CA)}. In general, {\em a CA is a dynamical system
which consists of a discrete-valued field defined at the sites of a regular
lattice and evolving in discrete time steps, with the field value at a site
being determined for the next time step by its present values in a
neighbourhood
of the site of interest} \cite{w}. In this work, particles reside on the sites
of a regular lattice; they move randomly between neighbouring sites, in
discrete
time steps, and react with a certain probability upon meeting. Formally, this
is
described by a set of {\em occupation numbers}, i.e.\ an integer-valued field
with a species label, giving the number of particles of the different species
at
each lattice site and evolving in time according to a local rule. The evolution
rule consists of a `transport' operation followed by a `reaction' operation.
The
two operations are repeated iteratively for consecutive time steps. Each
operation amounts to applying a simple algorithm on all sites of the lattice
simultaneously. The `transport' algorithm prescribes the probability with which
a particle will move to a neighbouring site; the probability may depend on
species label, location on the lattice, spatial direction or time; in the end,
each particle performs a random walk. The `reaction' algorithm, on the other
hand, provides the probability with which a given combination of occupation
numbers will lead to chemical reaction. We define macroscopic quantities by
averaging over an {\em ensemble} of independent copies of the system of
interest. {\em Particle density}, defined as an ensemble average of the
occupation number, obeys a discrete evolution equation. If chemical reactions
are present in this equation, occupation numbers can be entirely eliminated in
favour of locally averaged particle densities only if certain important
conditions are met. As will be  discussed in detail in Section \ref{model}, one
has to assume that the  chemical reactions do not give rise to correlations
among different species and  that the spatial dependence of macroscopic
quantities is sufficiently smooth.  The {\em continuum limit} of the resulting
equation (i.e.\ when lattice spacing  and time step go to zero) contains the
standard advection, diffusion and  reaction terms of the macroscopic PDE's
\footnote{For brevity we shall talk  about `diffusion', although, in the cases
of interest to us, it is {\em  mechanical dispersion}, rather than {\em
molecular diffusion}, that accounts  overwhelmingly for the {\em coefficient of
hydrodynamic dispersion}. To the  extent that mechanical dispersion can be
described by an effective diffusion  term, the two processes are
indistinguishable from the modelling point of view;  naturally, the
interpretation of the effective coefficient is very different in  the two cases
\cite{be}. We note incidentally that we find it sufficient for our purposes to
use the term `species' in a generic sense, although one  distinguishes in
principle between elementary {\em components} and composite {\em species}.}.
The
probabilities of motion and reaction can be chosen so that the  desired
macroscopic physical and chemical parameters are obtained.

The standard way to model reaction-transport phenomena consists in applying (i)
{\em local mass conservation} to derive a set of partial differential
equations,
and (ii) {\em local chemical equilibrium}, according to the law of mass action,
to derive a set of non-linear algebraic equations (assuming implicitly that
chemical equilibrium is attained instantaneously across the averaging volume on
the time scale of the transport processes). The ensuing system of coupled,
non-linear PDE's can be solved by a variety of numerical techniques. A common
problem is the difficulty in establishing general criteria guaranteeing the
{\em stability} of the numerical algorithms for solving systems of non-linear
PDE's; these questions are handled on a case by case basis. Concerning the
computer implementations of these solvers, the memory and time requirements
present an extremely difficult task for practical two- and three-dimensional
problems with present-day computers. As suggested in Ref.~\cite{FDE},
algorithms
that apply transport and chemical reactions sequentially appear to hold more
promise. We have already seen that this iterative aspect is shared by our
model.

The approach proposed in this work has been motivated by the following
arguments:

\begin{list}{(\alph{notes})}{\usecounter{notes}}

\item \refstepcounter{note} We model physicochemical processes at a level
intermediate between the {\em macroscopic level} described by PDE's
and the {\em microscopic level} of molecular dynamics. The `particles' of our
model are mathematical abstractions of the actual molecules of various chemical
species. The number of particles is large enough to make statistical concepts
meaningful, but is still many orders of magnitude smaller than the real
number of molecules involved. The CA is intended to describe macroscopic
behaviour which does not depend on the details of the microscopic dynamics,
but follows from general properties of the latter. The CA guarantees these
essential properties by applying them directly to its elementary constituents,
in a much more intuitive way than the continuum approach. For example, the
constraints of mass and momentum conservation, together with some smoothness
assumptions about the macroscopic variables, lead to the Navier-Stokes
equations
of fluid dynamics \footnote{Smoothness assumptions are indispensable for the
derivation of macroscopic PDE's, such as the Navier-Stokes equations, from
microscopic equations, such as the Boltzmann equation.}. Similarly, the
diffusion equation follows essentially from the random nature of collisions at
the microscopic level. The resulting computer code is simple and allows easy
variation of the dynamics and the boundary conditions. Moreover, statistical
fluctuations are inherently present in the CA, whereas PDE's rely on being able
to define a physical quantity, e.g. particle density, as a continuous variable
and thus neglect any local variations due to the microscopic nature of the
process. This difference is crucial when the fluctuations have significant
macroscopic consequences, as for example in the reaction $a+b \rightarrow
nothing \cite{tw}$.

\item By working with {\em integers}, the CA simulation is free of {\em
round-off errors}, which arise from the representation of real numbers by
finite computer words and lead to possible instabilities of the numerical
algorithms  used to solve PDE's. This is a significant advantage when
non-linear
reaction  terms are present, making the stability of PDE solvers hard to
establish. All quantities in a CA simulation are intrinsically finite and
infinities can  only arise by extrapolation to some limit (e.g.\ infinite
volume), in very much  the way this is done with real systems~\cite{t}. The
continuity required of the  solutions of PDE's is, by contrast, physically
inaccurate, as it  presupposes an infinitesimal limiting procedure (e.g.\
averaging over a volume  that shrinks to a point) which does not correspond to
reality below a minimal  scale (e.g.\ the volume per particle when evaluating
particle density).

\item CA algorithms are naturally parallel, i.e.\ they process a large number
of
data simultaneously. This makes them suitable for massively parallel computers.
In particular, the simple nature of the elementary physicochemical processes
should allow simulation on massively parallel computers with relatively
unsophisticated processors. Moreover, the local character of these processes
implies that only a minimal amount of communication between physically
neighbouring processors will be necessary. Computers with large numbers of
processors (up to tens of thousands) operating in parallel and favouring
next-neighbour communication are among those leading the push towards teraflop
performance ($10^{12}$ floating point operations per second) \cite{bo}.
Supercomputers are also being used to solve PDE's with standard numerical
methods, and the question arises if CA can provide a more efficient alternative
(when fluctuation effects are uninteresting). In this respect, we note that the
stochastic nature of the CA approach makes it in general slower than
deterministic methods of solving PDE's, if the same time and space
discretisation is used. On the other hand, these deterministic methods suffer
invariably from stability problems, which can be overcome, but usually at great
expense to the computation time. Here the CA approach has the important
advantage that it is inherently stable. We therefore expect that the answer to
the above question will depend highly on the specific application of interest.

\setcounter{notes}{0} \end{list}

The systems simulated in this work are conventionally described by
reaction-transport equations of the form derived in Section \ref{model} (Eq.\
(\ref{reac-tran-eqC})). As discussed above, these equations are based on
several
oversimplifying assumptions about the behaviour of real systems, but we shall
use them here  anyway as a reference point in testing our simulations. We
justify this choice  by the fact that reaction-transport PDE's lie
invariably at the basis of all known models for the systems in which we are
interested. We remain naturally  open to the possibility that discrepancies
between the CA approach and the solutions of the reaction-transport equations
may arise from some element of   microscopic reality (e.g.\ microscopic
fluctuations) that the continuum approach fails to capture. This may lead to
interesting corrections to the standard point of view. When no such corrections
arise, the CA may be used to approximate the  solutions of the continuum
equations; it is of course essential to study how  the result of the simulation
converges to the continuous function it is  approximating, as we refine the
discretisation of space and time. The merits of  the CA approach will then be
judged in comparison with other numerical methods.  {\em The aim of this work
is
to establish that the proposed CA can simulate a wide variety of
physicochemical
phenomena}, ranging from simple annihilation reactions to complex autocatalytic
reaction schemes leading to pattern formation. {\em The discrete approach will
be shown to be capable of approximating the  solution of the reaction-transport
PDE's. In some cases the results of the  discrete and continuum approaches will
disagree and in one of these the  discrepancy will point to a fundamental
shortcoming of the continuum approach.}

CA simulations are currently being performed by scientists in various
disciplines, who seek in the simplicity of their elementary algorithms the
unifying principles behind the often complex phenomena they observe \cite{CA}.
Thus, fluid motion is being modelled via {\em lattice-gas automata} (LGA).
Lattice-gas models have existed for several years and they were known to
display
certain hydrodynamic properties \cite{hpp}, but the recent increase in their
popularity \cite{d} has been largely brought about (i) by the advent of
powerful supercomputers, and (ii) by the realisation that they can approximate
hydrodynamic PDE's \cite{fhp}. LGA are special cases of deterministic CA. Our
approach is more closely related with probabilistic LGA models of
reaction-diffusion processes \cite{ldkb,cd}. All LGA models impose a limitation
on the number of particles per lattice site, in the form of an {\em exclusion
principle}. This is necessary when simulating CA on special purpose computers,
which can support a small number of bits per site (e.g.\ CAM-6, \cite{tm}), and
is useful for optimisation purposes on vector supercomputers. With computers
becoming available that perform floating-point operations on many processors in
parallel, we see no reason to maintain such a restriction, in particular since
the enforced small number of particles per site severely limits the statistics
of the simulation and makes either a bigger lattice or more simulations
necessary. In addition, probabilistic models of diffusion with an exclusion
principle are not able to model advection without introducing at the same time
unwanted non-linearities \cite{bl}. In our model, by contrast, advection
emerges
naturally from the microscopic rule. An exclusion principle also makes it
impossible to model local chemical reactions of arbitrary complexity.
Increasingly complex reactions can be brought into our model with minor
programming effort. Any model aiming at the description of the transport and
chemistry of complex solute systems in geological media \underline{must} be
able
to incorporate advection and arbitrary chemical reactions naturally. The model
proposed here appears, in this light, most suitable for applications to systems
of this kind.

The paper is organised as follows: In Section \ref{model} the discrete model is
introduced and the continuum limit is derived with special emphasis on the
physical assumptions that lead to the desired continuum equations.
Alternative microscopic rules and boundary conditions are also discussed.
Extensive simulations of physicochemical systems and detailed comparisons with
the corresponding differential equations are undertaken in Section \ref{simul}.
Discrepancies with the continuum approach will receive particular attention.
Finally, in Section \ref{C-O}, we discuss computational feasibility, summarise
our conclusions and set tasks for the future.

\newpage
\section{The Model} \label{model}
\setcounter{equation}{0}
\subsection{Time Evolution}
\indent

Our aim in this section is to formulate a microscopic model for the time
evolution of a system of particles moving randomly (with a possible bias in a
given direction) and reacting chemically among themselves. The particles can be
thought of as molecules in solution, migrating in a porous medium. The model
will be general enough to simultaneously describe the diffusion and advection
of several chemical species with different transport properties and subject to
various chemical reactions.

In our model, space and time are discrete: particles reside on the sites of a
regular lattice, with spacing $\lambda$, and the system evolves by a sequence
of
instantaneous transitions separated by time $\tau$. In each transition, the
system moves to a new state according to a local rule applied to all sites of
the lattice, and the procedure is iterated in discrete time steps. Below, we
shall define the local rule and proceed to show that, under certain conditions,
the equations describing the time evolution of the discrete system go over to a
set of differential equations in the limit $\lambda, \tau \rightarrow 0$.

We define an {\em evolution step} to consist of a {\em transport step} and a
{\em chemical reaction step}. During the transport step, we consider each
lattice site independently and, for each particle present, we make a random
decision whether it will move or stay stationary, with probabilities chosen so
as to obtain the desired macroscopic parameters. Having redistributed particles
this way, we proceed to the chemical reaction step, in which we decide for each
site independently whether the particles occupying it will react, with
probabilities reflecting the macroscopic reaction rates. The reaction
probability can be chosen according to a variety of rules, some of which may
give rise to the same macroscopic behaviour. A minimal prerequisite for a
reaction to take place is that there are sufficient particles of each reactant.
For the reactions that do go ahead, we remove particles of the reactant species
and add particles of the product species, according to the reaction equation.
Having completed the chemical reaction step we proceed to the next evolution
step and so on.

We begin the precise formulation of our model by considering a system of
particles belonging to various species $s_\alpha$ (for most purposes the index
$\alpha$ will be used interchangeably with the name $s_\alpha$ of the species).
The particles are located on the sites of a lattice $\cal L$. For simplicity,
we
consider a one-dimensional lattice. Extensions to higher dimensions are
straightforward. We focus on a typical site at position $x$ (we shall refer to
it for brevity as `site $x$') in the interior of the lattice. At time $t$,
there
are $N_{\alpha}(x,t)$ particles of species $\alpha$ at $x$. We refer to
$N_{\alpha}(x,t)$ as the occupation number. The values of $N_{\alpha}(x,t)$ for
all species $\alpha$ and lattice sites $x$ define the state of the system of
particles at time $t$. We assume a given distribution, $N_{\alpha}(x,0)$, of
particles at $t = 0$, which provides the initial condition to our problem, and
defer to a later point the discussion of boundary conditions. The evolution
during a time step $\tau$, from time $t$ to $t+\tau$, amounts to applying on
$N_\alpha(x,t)$ the product ${\cal C \circ T}$ of the operators describing
transport, $\cal T$, and chemical reactions, $\cal C$:
\begin{equation}
N_\alpha(x,t+\tau) = {\cal C \circ T} N_\alpha(x,t) .
\label{tran-reac-op}
\end{equation}

The purpose of the transport operation is to move particles by one lattice
spacing to the right with probability $p$, or to the left with probability $q$
$(p+q \leq 1)$. This is completed in two consecutive stages. Each stage
consists
of a local operation applied simultaneously to all lattice sites. Since the
action of $\cal T$ is defined for any set of occupation numbers, independent of
a particular state of the system, we introduce a generic set of occupation
numbers $\{ {\cal N}_\alpha(x): x \in {\cal L}-\partial {\cal L} \}$, where
$\partial {\cal L}$ represents the lattice boundary. First, for every lattice
site $x$ and every species $\alpha$, a random triplet $\mbox{\boldmath
$\xi$}_{x,n} \equiv ( \xi_{x,n}^{(-1)}, \xi_{x,n}^{(0)}, \xi_{x,n}^{(+1)} )$ is
drawn for each of the ${\cal N}_\alpha(x)$ particles of species $\alpha$
present
at $x$ ($n = 1,...,{\cal N}_\alpha(x)$). $\mbox{\boldmath $\xi$}_{x,n}$ takes
one of the values $(1,0,0), (0,1,0)$ or $(0,0,1)$ with probabilities $p, 1-p-q$
and $q$ respectively. All \mbox{\boldmath $\xi$}'s are drawn anew at each
update, but an explicit time index is omitted for simplicity. These triplets
are
stored for the second stage. We begin the second stage with a new,
\underline{empty} lattice and build the new occupation numbers according to
\begin{equation}
{\cal T} {\cal N}_\alpha(x) = \sum_{j=-1}^1 \sum_{n=1}^{\infty}
\xi_{x+j\lambda,n}^{(j)} \theta ( {\cal N}_\alpha (x+j\lambda) - n ),
\label{tran-op}
\end{equation}
where $j, n$ are integers and $\xi_{x+j\lambda,n}^{(-1)},
\xi_{x+j\lambda,n}^{(0)}, \xi_{x+j\lambda,n}^{(+1)}$ are the components of the
random triplet $\mbox{\boldmath $\xi$}_{x+j\lambda,n}$. The $\theta$-function
is
defined by $\theta(m) = 1$ for $m \geq 0$, and $\theta(m) = 0$ otherwise.

To show how the empty lattice is filled, we recall that ${\cal T}
{\cal N}_\alpha(x) = 0$ at the beginning of the second stage. The site $x$ is
fed with particles from its {\em immediate neighbourhood}, i.e.\ the sites
preceding and following $x$, as well as the site itself (in 2 dimensions, the
immediate neighbourhood of a site consists of 5 or 7 sites, for a square or
triangular lattice respectively). On the RHS of Eq.~(\ref{tran-op}), $j$ runs
over the sites of this neighbourhood, which are $x-\lambda, x$ and $x+\lambda$.
For each one of these sites and for each particle present at the same site of
the original lattice, we look up the corresponding random triplet stored
previously. If the component $\xi_{x+j\lambda,n}^{(j)}$ is 1, one particle is
added to ${\cal T} {\cal N}_\alpha(x)$, if it is 0, ${\cal T}
{\cal N}_\alpha(x)$ is left unmodified. In physical terms this corresponds to
particles moving to the right or left or remaining stationary with
probabilities
$p, q$ and $1-p-q$ respectively. The fact that one component of
$\mbox{\boldmath
$\xi$}_{x,n}$ is always 1 and the rest 0 implies that the transport operation
will place exactly one particle in the immediate neighbourhood of a site for
each of the particles originally at that site, thus conserving particle number.
Then Eq.~(\ref{tran-op}), with $N_\alpha(x,t)$ substituted for
${\cal N}_\alpha(x)$, can be interpreted as yielding the number of particles of
species $\alpha$ at $x$, after the transport phase of the update following time
$t$: particles from the left (i.e.\ on site $x-\lambda$ at time $t$) move to
$x$
with probability $p$, particles from the right (i.e.\ on site $x+\lambda$ at
time $t$) move to $x$ with probability $q$ and particles at $x$ remain there
with probability $1-p-q$. The last probability can be related to a {\em
retention factor} in the movement of the species $\alpha$ (see Section
\ref{simul}).

Next we define the chemical reaction operator $\cal C$ using the equation:
\begin{equation}
{\cal C} {\cal N}_\alpha(x) = {\cal N}_\alpha(x) + \sum_{r=1}^R
\left( \nu_{\alpha r}^{(f)} - \nu_{\alpha r}^{(i)} \right) \eta_{x,r}.
\label{reac-op}
\end{equation}
Here the summation runs over all chemical reactions in the problem at hand and
$\eta_{x,r}$ is a Boolean random variable. $\nu_{\alpha r}^{(i)}$ and
$\nu_{\alpha r}^{(f)}$ are the {\em stoichiometric coefficients} referring to
initial and final species of the reaction $r$ respectively. Naturally one can
talk about `initial' and `final' species only if $r$ is a one-way reaction. In
fact, most of the reactions we expect to encounter are reversible, of the type
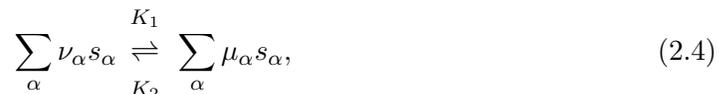
\begin{equation}
\sum_\alpha \nu_\alpha s_\alpha
\begin{array}{c} \mbox{\scriptsize $K_1$} \\ \rightleftharpoons \\
\mbox{\scriptsize $K_2$} \end{array}
\sum_\alpha \mu_\alpha s_\alpha,
\label{typ-rea}
\end{equation}
where $K_1, K_2$ are the {\em rate constants} and the stoichiometric
coefficients $\nu_\alpha, \mu_\alpha$ are non-negative integers. The summations
run over all species, but the stoichiometric coefficients vanish for species
not
involved in the reaction. It is convenient to express the above reaction as two
one-way reactions, namely
\begin{eqnarray}
\addtocounter{zaehler}{1}
\sum_\alpha \nu_\alpha s_\alpha
 & \begin{array}[b]{c} \mbox{\scriptsize $K_1$} \\ \rightarrow \end{array} &
\sum_\alpha \mu_\alpha s_\alpha, \\
\addtocounter{equation}{-1} \addtocounter{zaehler}{1}
\sum_\alpha \mu_\alpha s_\alpha
 & \begin{array}[b]{c} \mbox{\scriptsize $K_2$} \\ \rightarrow \end{array} &
\sum_\alpha \nu_\alpha s_\alpha.
\end{eqnarray}
\setcounter{zaehler}{0}
In Eq.\ (\ref{reac-op}) $R$ is the total number of one-way reactions thus
obtained and the index $r$ runs over all such reactions (if there are
irreversible  reactions, they are added to the list as they are). Thus, if the
first of the  above two reactions occupies the $r$-th position on the list of
one-way reactions, we identify $\nu_\alpha$ and $\mu_\alpha$ with $\nu_{\alpha
r}^{(i)}$ and $\nu_{\alpha r}^{(f)}$ respectively. In Eq.~(\ref{reac-op})
$\eta_{x,r}$ determines whether reaction $r$ takes place ($\eta_{x,r} = 1$) or
not ($\eta_{x,r} = 0$). In the latter case ${\cal N}_\alpha(x)$ is left
unmodified (as far as reaction $r$ is concerned), whereas in the former we
subtract the  number of reacting particles ($\nu_{\alpha r}^{(i)}$) and add the
number of  particles produced ($\nu_{\alpha r}^{(f)}$).

The probability with which $\eta_{x,r}$ is equal to 1 determines the rate at
which  reaction $r$ takes place and should therefore be chosen to reflect the
physical situation. In the physical case, a reaction rate depends both on
the rate constant ($K$) and on the concentration ($C_\alpha$) of reactants;
thus
for the reaction $a + 2b \rightarrow c$, the standard rate law gives the rate
of
production of $c$ as the product $K C_a C_b^2$. Therefore, the probability of
reaction has to depend on the rate constant and on the number of reactant
particles available at the site of interest. Whereas we treat the rate constant
as an intrinsic measurable property of the reaction, the functional dependence
of the microscopic rule on occupation numbers ought to yield the concentration
dependence of the standard rate law in the continuum limit. Thus, the
probability that reaction $r$ will take place at $x$ (i.e.\ for $\eta_{x,r} =
1$) can be written as
\begin{equation}
\wp (\eta_{x,r}=1 | \{ {\cal N}_\beta(x): s_\beta \in {\cal S} \} ) \equiv P_r
F_r ( \{ {\cal N}_\beta(x) \} ),          \label{reac-prob}
\end{equation}
where $\cal S$ is the set of all species. $P_r$ is a real number in the
interval
$[0,1]$ and will be related to $K_r$ later. $F_r$ is a function of the
occupation numbers and by varying it we can define various microscopic rules.

A simple choice for $F_r$ is a function that ensures the presence of sufficient
particles for the reaction to take place once at the microscopic level. One
occurrence of reaction $r$ means that $\nu_{\alpha r}^{(i)}$ $\alpha$-particles
are subtracted and $\nu_{\alpha r}^{(f)}$ are added at the particular site.
According to the rule defined with this $F_r$, {\em if there are sufficient
reactant particles present, the reaction will take place \underline{at most
once}, with probability $P_r$}. Formally we define $F_r$ as follows:
\begin{equation}
F_r ( \{ {\cal N}_\beta(x): s_\beta \in {\cal S} \} ) \equiv \prod_\beta
\theta \left( {\cal N}_\beta(x) - \nu_{\beta r}^{(i)} \right)
\hspace{1cm} \mbox{(Rule I)},      \label{F1}
\end{equation}
where the $\theta$-function guarantees that sufficient particles are present
and the product runs over all species. We recall the convention that
$\nu_{\beta r}^{(i)} = 0$ if $s_\beta$ is not a reactant in reaction $r$: if
that is the case, the contribution of species $\beta$ to the product is
trivially equal to one, since all ${\cal N}_\beta$'s are non-negative, so that
$F_r$ depends effectively only on the occupation numbers of the species
involved
as reactants in reaction $r$.

According to the rule defined by Eq.~(\ref{F1}), once there are sufficient
reactant particles for a reaction to take place, the latter proceeds with a
probability which does not further depend on the occupation numbers. Rule I
leads to the standard rate law in the continuum limit, but only for low
particle
densities. On the other hand we expect that, the greater the number of
particles
present, the likelier should be the reaction. We introduce therefore another
rule, in which the reaction probability is weighted by a product involving the
number of particles of each species. We allow, namely, species $\beta$ to
contribute a factor $\prod_{m=1}^{\nu_{\beta r}^{(i)}}({\cal N}_\beta(x) - m +
1 )$ to $F_r$. {\em The principal motivation for this choice is that it leads
to
the standard rate law for any particle density}. Thus, the new rule amounts to
defining
\begin{equation}
F_r( \{ {\cal N}_\beta(x): s_\beta \in {\cal S} \} ) =
\prod_\beta\prod_{m=1}^{\nu_{\beta r}^{(i)}}({\cal N}_\beta(x) - m + 1 )
\hspace{1cm} \mbox{(Rule II)}.      \label{F2}
\end{equation}
Here we allow the product to run over all species, with the convention
$\prod_{m=1}^{0} ( {\cal N}_\beta(x) - m + 1 ) = 1$, which ensures that species
$\beta$ will contribute a factor 1 if it is not a reactant in reaction $r$
($\nu_{\beta r}^{(i)} = 0$).

Eqs.~(\ref{tran-reac-op}), (\ref{tran-op}), (\ref{reac-op}), (\ref{reac-prob})
and (\ref{F1}) or (\ref{F2}) define the rule of evolution of the system of
interacting particles. In the following subsections we are going to use these
equations in order to derive discrete evolution equations, analogous to the
finite difference equations (FDE's) approximating the differential
reaction-transport equations. For this purpose we need to define a real-valued
field, the {\em particle density} $\rho_\alpha(x,t)$, from the integer-valued
occupation number $N_\alpha(x,t)$. It is clear of course that only
$N_\alpha(x,t)$ is involved in the simulation, whereas $\rho_\alpha(x,t)$ can
be
optionally defined at the level of the output and is not fed back into the
simulation.

The occupation numbers obtained in a simulation of the discrete model possess
a granularity which is not manifest in a typical measurement. Measured
quantities are effectively averaged over a volume depending on the spatial
resolution of the measuring apparatus. It is anyway clear that physical
attributes of extended systems can only be defined after averaging over a
certain minimal length scale. The continuum approach appears to describe
physical quantities over arbitrarily short distances, but this becomes
meaningless below the above minimal scale. Our approach lies closer to reality
in that the transition from integers (occupation numbers) to real numbers
(particle densities~\footnote{We reserve the concept of {\em concentrations}
for the physically measurable quantities, which will be obtained later by
multiplying the {\em densities} by a scaling factor.}) is effected via an
averaging procedure. This can be performed either over a neighbourhood of
lattice sites or over corresponding sites in an {\em ensemble} of states
obtained by several independent simulations of the system. We make a {\em
smoothness assumption}, according to which the particle density does not vary
appreciably on the length scale of the neighbourhood used in the first of the
above averaging procedures. The average densities of two such neighbourhoods
may of course differ appreciably if their separation is on a larger scale.
Under the smoothness assumption, the results of the above two averaging
procedures should agree, within fluctuations. Based on this discussion, we
define the particle density
\begin{equation}
\rho_\alpha (x,t) \equiv < N_\alpha(x,t) >,  \label{den-def}
\end{equation}
where $< \ldots >$ denotes an {\em ensemble average}.

\subsection{Transport Equation} \label{transport}
\indent

{}From a numerical point of view, transport differs from chemical reactions in
two
important respects: (i) The former involves communication of information
between
neighbouring sites; in the continuum limit this gives rise to derivative terms
(advection, diffusion), which depend on the precise way the limit is taken.
(ii)
In our model, the evolution equations for pure transport are linear in the
density and the respective FDE's are subject to well-defined stability
criteria;
this distinguishes them from the reaction-transport equations, which are
usually
non-linear and follow no general stability criteria. For these reasons it will
be worthwhile to concentrate first on pure transport. We begin, therefore, by
replacing ${\cal C}$ in Eq.~(\ref{tran-reac-op}) by the identity operator. Then
the species propagate independently of each other and we can drop, for the
purposes of the present derivation, the index $\alpha$.

Combining Eqs.~(\ref{den-def}) and (\ref{tran-reac-op}), the latter with
${\cal C} \equiv 1\hspace{-1.5mm}1$, we obtain
\begin{equation}
\rho(x,t+\tau ) = < N(x,t+\tau) > = < {\cal T} N(x,t) >.   \label{tran-only}
\end{equation}
According to Eq.~(\ref{tran-op}),
\begin{eqnarray}
< {\cal T} N(x,t) > & = & \sum_{n=1}^{\infty} \left(
< \xi_{x-\lambda,n}^{(-1)} > < \theta ( N(x-\lambda,t) - n ) > +
< \xi_{x,n}^{(0)} > < \theta ( N(x,t) - n ) > \right. \nonumber \\ & & +
\left. < \xi_{x+\lambda,n}^{(+1)} > < \theta ( N(x+\lambda,t) - n ) > \right)
\nonumber \\
 & = & p < N(x-\lambda,t) > + (1-p-q) < N(x,t) > + q
< N(x+\lambda,t) > \nonumber \\
 & = & p\rho(x-\lambda,t) + (1-p-q)\rho(x,t) + q\rho(x+\lambda,t) .
\label{tran-av}
\end{eqnarray}
Eq.~(\ref{tran-av}) is exact and follows from the fact that the \mbox{\boldmath
$\xi$}'s are statistically independent of the $N$'s. In deriving
(\ref{tran-av})
we made use of the expectation values of the components
$\xi^{(j)}_{x,n}$:
\begin{equation}
\left( \begin{array}{c} < \xi_{x,n}^{(-1)} > \\ < \xi_{x,n}^{(0)}> \\
< \xi_{x,n}^{(+1)} > \end{array} \right) = < \mbox{\boldmath $\xi$}_{x,n} > =
p \left( \begin{array}{c} 1 \\ 0 \\ 0 \end{array} \right) +
(1-p-q) \left( \begin{array}{c} 0 \\ 1 \\ 0 \end{array} \right) +
q \left( \begin{array}{c} 0 \\ 0 \\ 1 \end{array} \right) =
\left( \begin{array}{c} p \\ 1-p-q \\ q \end{array} \right) ,
\end{equation}
as well as of the simple result
\begin{eqnarray}
\sum_{n=1}^{\infty} < \theta ( N(x,t) - n ) > & = &
< \sum_{n=1}^{\infty} \theta ( N(x,t) - n ) > \nonumber \\
& = & < N(x,t) > \hspace{1cm}, \forall x \in {\cal L} .        \label{sum-th}
\end{eqnarray}
The last equality follows from the fact that the $\theta$-function contributes
1
to the sum for each value of $n$ between 1 and $N(x,t)$ and 0 for higher
values.
We shall show the same result in a different way later
(Eq.~(\ref{sum-prob-geq})).

{}From Eqs.~(\ref{tran-only}) and (\ref{tran-av}) we deduce the evolution
equation
\begin{equation}
\rho(x,t+\tau ) = p\rho(x-\lambda,t) + (1-p-q)\rho(x,t) + q\rho(x+\lambda,t) .
\label{evol-tran}
\end{equation}
Eq.~(\ref{evol-tran}) can be readily rearranged as follows
\begin{eqnarray}
\lefteqn{ {\rho(x,t+\tau) - \rho(x,t) \over \tau} } \nonumber \\
 & & = - V \, {\rho(x+\lambda,t) - \rho(x-\lambda,t) \over 2\lambda} +
  D \, {\rho(x+\lambda,t) - 2\rho(x,t) + \rho(x-\lambda,t) \over \lambda^2},
\label{FD-tran}
\end{eqnarray}
where we have defined
\begin{equation}
V \equiv (p-q){\lambda \over \tau}      \hspace{1cm} , \hspace{1cm}
D \equiv (p+q){\lambda^2 \over 2\tau} \hspace{0.25cm} . \label{VD}
\end{equation}

In Eq.~(\ref{FD-tran}) we recognise the forward difference approximation for
the first time derivative and the central difference approximations for the
first and second space derivatives of the density:
\begin{eqnarray}
{\rho(x,t+\tau) - \rho(x,t) \over \tau} & = &
{\partial\rho(x,t) \over \partial t} + O(\tau) , \nonumber \\
{\rho(x+\lambda,t) - \rho(x-\lambda,t) \over 2\lambda} & = &
{\partial\rho(x,t) \over \partial x} + O(\lambda^2) , \label{taylor} \\
{\rho(x+\lambda,t) - 2\rho(x,t) + \rho(x-\lambda,t) \over \lambda^2} & = &
{\partial^2\rho(x,t) \over \partial x^2} + O(\lambda^2). \nonumber
\end{eqnarray}
Substituting these approximations in Eq.~(\ref{FD-tran}) we arrive at the
equation
\begin{equation}
\frac{\partial\rho(x,t)}{\partial t} + O(\tau) =
- V \frac{\partial\rho(x,t)}{\partial x}
+ D \frac{\partial^2\rho(x,t)}{\partial x^2} + O(\lambda^2).
\label{almost-tran-eq}
\end{equation}
Taking the limit $\lambda \rightarrow 0$, $\tau \rightarrow 0$ and $p-q
\rightarrow 0$ in such a way that $\lambda^2 /\tau$ and $(p-q)\lambda /\tau$
remain finite, we obtain the transport equation
\begin{equation}
\frac{\partial\rho(x,t)}{\partial t} = -V \frac{\partial\rho(x,t)}{\partial x}
+D \frac{\partial^2\rho(x,t)}{\partial x^2} \hspace{0.25cm} . \label{tran-eq}
\end{equation}
On the RHS of Eq.~(\ref{tran-eq}) there is an advective term, with velocity
$V$, and a diffusive/dispersive term, with {\em diffusion coefficient} $D$. We
note that the continuum limit is taken in such a way that $V$ and $D$,
as defined in Eq.\ (\ref{VD}), remain finite. Eq.~(\ref{FD-tran}) is the
forward-time centred-space finite difference  approximation to the transport
equation (\ref{tran-eq}). We have thus shown  that, {\em as far as pure
transport is concerned, our model constitutes a stochastic way of solving the
finite difference approximation to the continuum  PDE}.

A few remarks are in order here:
\begin{list}{(\alph{notes})}{\usecounter{notes}}

\item \refstepcounter{note} The differential equation obtained in the continuum
limit depends on the way the limit is approached. Let us first try to keep
$p-q$
finite. We consider the migration of an ensemble of particles concentrated
initially at the same lattice site. The density of particles evolves according
to Eq.~(\ref{FD-tran}), whose RHS contains a term proportional to $p - q$
(`advection') and one proportional to $p + q$ (`diffusion'). After a finite
time
$t \gg \tau$, diffusion would result in an average displacement $\sim
\sqrt{\lambda^2 (t /\tau)}$, whereas the particles would propagate a distance
$\sim \lambda (t /\tau)$ in the same time due to advection. As $\lambda
\rightarrow 0$ and $\tau \rightarrow 0$, but keeping $\lambda/\tau$ constant,
an infinite number of updates (given by $t / \tau$) would be needed to make the
latter distance finite, but it is after an infinitely larger number of steps
(given by $(t / \tau)^2$) that the former displacement would have a chance to
become finite. In other words, diffusion would become infinitely slower than
advection and the ensemble of particles would move without spreading. Thus, if
we keep $p-q$ and $\lambda /\tau$ finite, then only the advective term survives
as $\lambda, \tau \rightarrow 0$. If on the other hand one tries to keep $p-q$
and $\lambda^2 / \tau$ finite, advection becomes  infinitely fast and this
particular limit is of no practical interest. The two processes will be
comparable in the continuum limit, however, if $p-q$ vanishes like $\lambda$.
$p-q$ is the bias in right/left displacement and gives the size of the
advective
velocity in units of lattice spacings per time step. By letting $p-q
\rightarrow
0$ we curb the uncontrolled growth of the advective  displacement, while
preserving a finite average diffusive spread.

\item From the definitions of $V$ and $D$ and the obvious conditions $p-q \leq
p+q \leq 1$, we deduce
\begin{equation}
V {\tau \over \lambda} \leq 2 D {\tau \over \lambda^2} \leq 1 \hspace{0.25cm} .
\label{stab}
\end{equation}
In Eq.~(\ref{stab}) we recognise the stability conditions for the FDE
(\ref{FD-tran}) \cite[Eq.~(5.1.18)]{noye}. These stability conditions are
imposed by hand in the usual FDE approach to ensure that the round-off error
introduced by numerical computation does not increase exponentially. Since we
have shown the equivalence of our method to the FDE, it is not surprising that
these inequalities hold, but it is an important feature of the random walk that
they are implemented in an automatic and natural fashion.

\item If $p+q = 1$, the system is subject to the so-called `checkerboard
parity' \cite{lever}. Thus, if we colour sites in a checkerboard fashion, two
particles occupying at one time sites of different colours will always be on
differently coloured sites and will never meet. The system divides into two
subsystems which alternate between sublattices of different colour, but remain
forever decoupled. We shall show later in this section that, when particles are
placed on a lattice according to a uniform random distribution, the occupation
numbers obey a Poisson distribution (Eq.~(\ref{prob-tran-n-also})).
The existence of two decoupled sublattices does not  influence this result, as
long as the particles are initially distributed evenly between the two
sublattices. The opposite extreme would be to place all  particles initially on
sites of one sublattice and then perform a large number  of diffusion steps. In
that case, one of the two sublattices is alternately  empty and the occupation
numbers end up satisfying the Poisson distribution as if all particles were
uniformly distributed only on the occupied sublattice: in
Eq.~(\ref{prob-tran-n-also}) the density has to be doubled (or, equivalently,
evaluated by dividing through half the number of lattice sites) and the
probability for occupation number 0 has to be augmented by a probability of 1/2
that the site belongs to the empty sublattice. In intermediate cases, in which
particles are placed with a bias favouring one sublattice, there will be a
corresponding deviation from the Poisson distribution, if the latter is
calculated on the assumption that particles are distributed over the entire
lattice. The symmetry just described will obviously not hold if $p+q < 1$,
i.e.\
if particles have a non-zero probability to remain stationary and thus populate
the same sublattice on successive time steps. Checkerboard parity can also
break
down because of boundary conditions. Thus, the subsystems may mix if we impose
periodic boundary conditions \cite{ldkb}.

\item Species with different transport properties (advection velocity or
diffusion coefficient) can be described on the same lattice by (i) moving
particles belonging to different species by different numbers of lattice
spacings at each update, or (ii) by moving species at different multiples of a
time step. This way we can simulate, for instance, a problem of pure diffusion,
where the diffusion coefficients of different species are in the
ratio of integers. A more natural and physically appealing way to describe
species with transport coefficients whose ratio may vary continuously is to
make
$p$ and $q$ species dependent. Given $V_\alpha$ and $D_\alpha$, we evaluate
$p_\alpha$ and $q_\alpha$ from the equations
\begin{equation}
p_\alpha = D_\alpha \frac{\tau}{\lambda^2} + \frac{V_\alpha}{2}
\frac{\tau}{\lambda} \hspace{0.5cm} , \hspace{0.5cm}
q_\alpha = D_\alpha \frac{\tau}{\lambda^2} - \frac{V_\alpha}{2}
\frac{\tau}{\lambda} \hspace{0.25cm} .         \label{pq}
\end{equation}
If we choose $\tau$ and $\lambda$ so that (i) $p_\alpha+q_\alpha = 1$ for the
species with largest $D_\alpha$ \footnote{In principle $p_\alpha+q_\alpha < 1$
for the species with largest $D_\alpha$ will do as well. This choice, however,
slows down the simulation and is not necessary, unless the value of $\tau$ has
to be small for reasons such as those explained at the end of Subsection
\ref{chemistry}.} and (ii) $\left| V_\alpha \right| \lambda/2D_\alpha \leq~1$
for the species with largest $\left| V_\alpha \right| /D_\alpha$, then the
conditions $p_\alpha \geq 0$, $q_\alpha \geq 0$ and $p_\alpha + q_\alpha \leq
1$
will be fulfilled for all species. Condition (i) determines $\tau$ as a
function
of $\lambda$ and condition (ii) makes sure that $\lambda$ is small enough to
make the first term on the RHS of the  equations (\ref{pq}) larger in absolute
value than the second. We note in passing that one might think of treating
cases
with inhomogeneous transport parameters by making the probabilities position
dependent. In that case the evolution equation (\ref{evol-tran}) becomes
\begin{equation}
\rho(x,t+\tau ) = p(x-\lambda)\rho(x-\lambda,t) +
\left[ 1-p(x)-q(x) \right] \rho(x,t) + q(x+\lambda) \rho(x+\lambda,t) ,
\label{evol-tran2}
\end{equation}
where $p$ and $q$ are functions of position and the species label need not be
explicitly indicated. From Eq.~(\ref{evol-tran2}) we readily derive the
transport equation
\begin{equation}
\frac{\partial\rho(x,t)}{\partial t} = - \frac{\partial}{\partial x}
\left[ V(x)\rho(x,t) \right] +
\frac{\partial^2}{\partial x^2} \left[ D(x)\rho(x,t) \right] , \label{tran-eq1}
\end{equation}
where $V(x)$ and $D(x)$ are related to $p(x)$ and $q(x)$ according to
Eq.~(\ref{VD}). Note that Eq.~(\ref{tran-eq1}) can be written in the more
familiar form \cite{be}
\begin{equation}
\frac{\partial\rho(x,t)}{\partial t} = - \frac{\partial}{\partial x}
\left[ W(x)\rho(x,t) \right] +
\frac{\partial}{\partial x} \left[ D(x){\partial\rho(x,t) \over \partial x}
\right]
\end{equation}
by defining $W(x) \equiv V(x) - dD(x)/dx$.

\item If we repeat the above derivation for a rectangular lattice in $d$
dimensions, assuming the same coefficient of diffusion in all directions, we
arrive at the result $D = P\lambda^2 / 2\tau d$, where $P$ is the probability
that a particle moves to another site during a transport step ($P = p + q$ in
one dimension). $P$ is a measure of particle mobility and is inversely
proportional to the {\em retention factor} $\cal R$ that will be introduced in
Section~\ref{simul}, in connection with sorbing species.

\setcounter{notes}{0} \end{list}

\subsection{Chemical Kinetics} \label{chemistry}
\indent

We now consider the full problem in which an evolution step is completed by the
action of the chemical reaction operator $\cal C$ on the result of the
transport
operation. From Eqs.\ (\ref{tran-reac-op}) and (\ref{reac-op}) we obtain:
\begin{eqnarray}
N_\alpha(x,t+\tau) & = & {\cal C \circ T} N_\alpha(x,t) \nonumber \\
                     & = & {\cal T} N_\alpha(x,t) + \sum_{r=1}^{R}
    \left( \nu_{\alpha r}^{(f)} - \nu_{\alpha r}^{(i)} \right) \eta_{x,r}
\label{more-tran-reac-op}
\end{eqnarray}
and
\begin{eqnarray}
\rho_\alpha(x,t+\tau) & = & < N_\alpha(x,t+\tau) > \nonumber \\
& = & < {\cal T} N_\alpha(x,t) > +
\sum_{r=1}^{R} \left( \nu_{\alpha r}^{(f)} - \nu_{\alpha r}^{(i)} \right)
    P_r < F_r( \{ {\cal T} N_\beta(x,t): s_\beta \in {\cal S} \} ) > ,
    \nonumber \\
 & & \mbox{} \label{reac-also}
\end{eqnarray}
where we have directly substituted for $< \eta_{x,r} >$ the expression derived
from Eq.\ (\ref{reac-prob}).

At this point we need a specific ansatz for $F_r$ in order to proceed further.
Assuming the form given in (\ref{F1}) or (\ref{F2}), we are faced with the
problem of taking the expectation value of a product of random variables. The
simplest possibility is that these variables are {\em mutually independent}.
This will be true if the occupation numbers of different species are mutually
independent; in other words, if the number of particles of every species at a
site is not correlated with the numbers of particles of the other species at
that site. Correlations can arise as a consequence of interactions among the
particles, i.e.\ collisions or reactions of all kinds. In our model, particles
do not explicitly collide, but are scattered by a random background (we can
think of each random scattering of a particle as simulating several collisions
of a solute molecule with solvent or other solute molecules). Therefore,
correlations can only arise as a result of chemical reactions. We shall see in
Section \ref{simul} that correlations do occur in our model. We
\underline{postulate}, nevertheless, {\em molecular chaos}, i.e.\ the absence
of correlations, for the purposes of our derivation and return to this point in
the next section. Then the average of the product equals the product of the
averages of the individual terms and we have to evaluate expressions of the
type $\left< \theta \left( {\cal T} N_\alpha(x,t) -  \nu_{\alpha r}^{(i)}
\right) \right>$ and $< \prod_{m=1}^{\nu_{\alpha r}^{(i)}} ( {\cal T}
N_\alpha(x,t) - m + 1 ) >$, for rules I and II respectively.

We first derive the evolution equations using reaction rule I (Eq.~(\ref{F1})).
By definition, $\theta \left( {\cal T}N_\alpha(x,t) \right.$ $\left. -
\nu_{\alpha r}^{(i)} \right)$ is equal to 1 when there are not less than
$\nu_{\alpha r}^{(i)}$ $\alpha$-particles at site $x$ after the transport
operation and 0 otherwise. It follows that the ensemble average of the
$\theta$-function gives the fraction of ensemble members for which there are
not
less than $\nu_{\alpha r}^{(i)}$ $\alpha$-particles at $x$. We define this
fraction to be the probability that there are sufficient $\alpha$-particles at
that position:
\begin{equation}
\wp \left( {\cal T} N_\alpha(x,t) \geq \nu_{\alpha r}^{(i)} \right) \equiv
\left< \theta \left( {\cal T} N_\alpha(x,t) - \nu_{\alpha r}^{(i)} \right)
\right>.                                     \label{prob-T-geq}
\end{equation}
$\{ N_\alpha(x,t): s_\alpha \in {\cal S} \}$ is as much a set of occupation
numbers as $\{ {\cal T} N_\alpha(x,t): s_\alpha \in {\cal S} \}$ and we can
temporarily neglect ${\cal T}$:
\begin{equation}
\wp ( N_\alpha(x,t) \geq n ) \equiv
\left< \theta \left( N_\alpha(x,t) - n \right) \right>,   \label{prob-geq}
\end{equation}
where $n$ is any non-negative integer. It is easy to see that
\begin{eqnarray}
\sum_{n=1}^{\infty} \wp ( N_\alpha(x,t) \geq n ) & = &
\sum_{n=1}^{\infty} \sum_{m=n}^{\infty} \wp ( N_\alpha(x,t) = m ) \nonumber \\
& = & \sum_{m=1}^{\infty} m \wp ( N_\alpha(x,t) = m ) \nonumber \\
& = & < N_\alpha(x,t) >.                                \label{sum-prob-geq}
\end{eqnarray}
We note that Eq.\ (\ref{sum-th}) follows from Eq.\ (\ref{sum-prob-geq})
and the definition (\ref{prob-geq}).

We now need to relate the probability $\wp ( N_\alpha(x,t) = n )$ to the
density
$\rho_\alpha(x,t)$. We are able to fulfill this task under the smoothness
assumption, which guarantees that our definition of the density as an ensemble
average ( Eq.\ (\ref{den-def}) ) is equivalent to the alternative definition as
a local spatial average. We employ, for the purpose of the present argument,
the
latter definition and recall that, according to the discussion above
Eq.~(\ref{den-def}), local averages are evaluated over sections of the lattice
which are inhabited by essentially homogeneous populations of particles.
Following Ref.~\cite{b}, let such a subsystem contain, at time $t$, $N$
$\alpha$-particles distributed evenly on $M$ lattice sites. If site $x$ belongs
to the particular section of the lattice, then $\rho_\alpha(x,t) \simeq N/M$
and
$\wp ( N_\alpha(x,t) = n )$ can only depend on $n$ and $\rho_\alpha(x,t)$. We
wish to calculate the probability $\wp ( N_\alpha(x,t) = n )$ that $n$
particles
(out of $N$) are found at $x$ ($M, N \gg n$). There are $\left(
\begin{array}{c} N \\ n \end{array} \right)$ ways of selecting $n$ particles
out
of $N$. Once the $n$ particles are placed at site $x$, the remaining $N-n$
particles can distribute themselves in $(M-1)^{N-n}$ ways on the remaining
$M-1$
sites. Thus there is a total of $\left( \begin{array}{c} N \\ n \end{array}
\right) (M-1)^{N-n}$ configurations with $n$ particles at $x$. The desired
probability is obtained by dividing the number of these configurations by the
number $M^N$ of all possible configurations of $N$ particles on $M$ sites:
\begin{eqnarray}
\wp ( N_\alpha(x,t) = n ) & = & \left( \begin{array}{c} N \\ n \end{array}
\right)
{(M-1)^{N-n} \over M^N} = {N! \over (N-n)! \; n!} {(M-1)^{N-n} \over M^N}
\nonumber \\
& \simeq & \frac{\rho_\alpha^n(x,t)}{n!} e^{-\rho_\alpha(x,t)}
\hspace{1cm} \mbox{for} \; \; n \ll M, N
\label{prob-tran-n-also}
\end{eqnarray}
which is just the Poisson distribution.
Adopting, for convenience, the compact notation $\rho_\alpha(x,t+\tau/2)
\equiv  < {\cal T} N_\alpha(x,t) >$, we write by analogy
\begin{equation}
\wp ( {\cal T} N_\alpha(x,t) = n ) \simeq
\frac{\rho_\alpha^n(x,t+\tau/2)}{n!} e^{-\rho_\alpha(x,t+\tau/2)}.
\label{prob-tran-n}
\end{equation}
No overdue significance should be attached to the fraction in the time argument
$t+\tau/2$: it merely denotes an intermediate situation just before execution
of the chemical reaction step.

Substituting the ansatz (\ref{F1}) in Eq.\ (\ref{reac-also}), applying the
factorisation of the expectation value that follows from the molecular chaos
hypothesis, employing the definition (\ref{prob-T-geq}) and finally
\underline{assuming an equality} for Eq.\ (\ref{prob-tran-n}), we find
\begin{eqnarray}
\rho_\alpha(x,t+\tau) & = & \rho_\alpha(x,t+\tau/2) +
\sum_{r=1}^{R} \left( \nu_{\alpha r}^{(f)} - \nu_{\alpha r}^{(i)} \right) P_r
\prod_\beta
\wp\left( {\cal T} N_\beta(x,t) \geq \nu_{\beta r}^{(i)} \right)
\nonumber \\
& = & \rho_\alpha(x,t+\tau/2) + \sum_{r=1}^{R}
\left( \nu_{\alpha r}^{(f)} - \nu_{\alpha r}^{(i)} \right) P_r \prod_\beta
\sum_{n=\nu_{\beta r}^{(i)}}^{\infty} \wp \left( {\cal T} N_\beta(x,t) = n
\right) \nonumber \\
& = & \rho_\alpha(x,t+\tau/2) + \sum_{r=1}^{R}
\left( \nu_{\alpha r}^{(f)} - \nu_{\alpha r}^{(i)} \right) P_r \prod_\beta
\sum_{n=\nu_{\beta r}^{(i)}}^{\infty} \frac{\rho_\beta^{n}(x,t+\tau/2)}{n!}
e^{-\rho_\beta(x,t+\tau/2)},     \nonumber \\
\mbox{} \label{evol-full}
\end{eqnarray}
where
\begin{equation}
\rho_\alpha(x,t+\tau/2 ) = p_\alpha\rho_\alpha(x-\lambda,t) +
               (1-p_\alpha-q_\alpha)\rho_\alpha(x,t) +
                             q_\alpha\rho_\alpha(x+\lambda,t) .
\label{evol-tran1}
\end{equation}

Rearranging terms as in Eq.\ (\ref{FD-tran}) we deduce from Eqs.\
(\ref{evol-full}) and (\ref{evol-tran1})
\begin{eqnarray}
\lefteqn{ {\rho_\alpha(x,t+\tau) - \rho_\alpha(x,t) \over \tau}}
\nonumber \\
 & = & - V_\alpha \,
  {\rho_\alpha(x+\lambda,t) - \rho_\alpha(x-\lambda,t) \over 2\lambda} +
  D_\alpha \,
  {\rho_\alpha(x+\lambda,t) - 2\rho_\alpha(x,t) +
                    \rho_\alpha(x-\lambda,t) \over \lambda^2}   \nonumber \\
 & & + \sum_{r=1}^{R}
\left( \nu_{\alpha r}^{(f)} - \nu_{\alpha r}^{(i)} \right)
k_r \prod_\beta \sum_{n=\nu_{\beta r}^{(i)}}^{\infty}
\frac{\nu_{\beta r}^{(i)}!}{n!} \rho_\beta^{n}(x,t+\tau/2)
e^{-\rho_\beta(x,t+\tau/2)} ,
\label{FD-full}
\end{eqnarray}
where
\begin{equation}
V_\alpha \equiv (p_\alpha-q_\alpha){\lambda \over \tau}   \hspace{1cm} ,
\hspace{1cm}
D_\alpha \equiv (p_\alpha+q_\alpha){\lambda^2 \over 2\tau}  \label{VD1}
\end{equation}
and the rate constant $k_r$ is defined as
\begin{equation}
k_r \equiv \frac{P_r}{\tau} \frac{1}{\prod_\alpha\nu_{\alpha r}^{(i)}!}
\hspace{1cm} \mbox{(Rule I)}.
\label{kr1} \end{equation}
$k_r$ should not be confused with the \underline{physical} rate constant $K_r$
which will be defined shortly.

This is as far as we can go with the discrete model. We now wish to derive the
continuum limit of the full model. With the continuum approximations
(\ref{taylor}), Eq.\ (\ref{FD-full}) becomes
\begin{eqnarray}
\frac{\partial\rho_\alpha(x,t)}{\partial t} & + & O\left(\tau\right)
= - V_\alpha \frac{\partial \rho_\alpha(x,t)}{\partial x} +
  D_\alpha \frac{\partial^2\rho_\alpha(x,t)}{\partial x^2} + O \left( \lambda^2
  \right) \nonumber \\
 & + & \sum_{r=1}^{R} \left( \nu_{\alpha r}^{(f)} - \nu_{\alpha r}^{(i)}
\right)
k_r \prod_\beta \sum_{n=\nu_{\beta r}^{(i)}}^{\infty}
\frac{\nu_{\beta r}^{(i)}!}{n!} \rho_\beta^{n}(x,t) e^{-\rho_\beta(x,t)} +
O(\tau).                \label{taylor1}
\end{eqnarray}

In Subsection \ref{transport} we let $\lambda, \tau$ and $p_\alpha-q_\alpha
\rightarrow 0$, while keeping $\lambda^2/\tau$ and $(p_\alpha-q_\alpha)
\lambda/\tau$ finite. In the presence of chemical reactions, we also take $P_r
\rightarrow 0$, but keep $P_r/\tau$ finite for all $r$. It follows
immediately that $p_\alpha-q_\alpha$ and $P_r$ are  $O(\lambda)$ and $O(\tau)$
respectively. Taking the  continuum limit of Eq.\ (\ref{taylor1}) as described,
we derive the  reaction-transport equation
\begin{equation}
\frac{\partial\rho_\alpha(x,t)}{\partial t} =
- V_\alpha \frac{\partial\rho_\alpha(x,t)}{\partial x}
+ D_\alpha \frac{\partial^2\rho_\alpha(x,t)}{\partial x^2} +
\sum_{r=1}^{R} \left( \nu_{\alpha r}^{(f)} - \nu_{\alpha r}^{(i)} \right) k_r
\prod_\beta \sum_{n=\nu_{\beta r}^{(i)}}^{\infty}
\frac{\nu_{\beta r}^{(i)}!}{n!} \rho_\beta^{n}(x,t) e^{-\rho_\beta(x,t)}
\hspace{.5cm} \mbox{(Rule I)}. \label{reac-tran-eq1}
\end{equation}

We note that the forward-time centred-space FDE corresponding to
Eq.~(\ref{reac-tran-eq1}) is similar but not identical to the discrete
evolution
equation (\ref{FD-full}). The difference lies in the time at which $\rho_\beta$
is evaluated: for the FDE one uses the value of $\rho_\beta$ at time $t$, while
for the evolution equation $\rho_\beta$ is evaluated at the end of the
transport
step performed at $t$. Moreover, iteration of Eq.~(\ref{FD-full}) involves a
two-step process where $\rho_\beta(x,t+\tau/2)$ is first generated through
Eq.~(\ref{evol-tran1}).

To obtain the standard form of the rate terms we have to consider low particle
densities ($\rho_\alpha \ll 1$). If the latter are sufficiently low, then the
sums in Eq.\ (\ref{reac-tran-eq1}) are dominated by the lowest-order terms and
we
obtain products of the type
$\prod_\beta \rho_\beta^{\nu_{\beta r}^{(i)}}$:
\begin{equation}
\frac{\partial\rho_\alpha(x,t)}{\partial t} \simeq
- V_\alpha \frac{\partial\rho_\alpha(x,t)}{\partial x}
+ D_\alpha \frac{\partial^2\rho_\alpha(x,t)}{\partial x^2} +
\sum_{r=1}^{R} \left( \nu_{\alpha r}^{(f)} - \nu_{\alpha r}^{(i)} \right) k_r
\prod_\beta \rho_\beta^{\nu_{\beta r}^{(i)}}(x,t) .
\label{reac-tran-eq-ld}
\end{equation}

Eq.\ (\ref{reac-tran-eq-ld}) is typical of the kind of equations numerically
solved by conventional approaches. It is therefore desirable that the discrete
model reduce to a set of such equations in some limit, under well-defined
assumptions. If that is the case, it is sensible to test simulations of the
discrete model by making sure that their predictions  converge to the solutions
of the appropriate differential equations. The chemical reaction rule defined
by
Eqs.\ (\ref{reac-op}), (\ref{reac-prob})  and (\ref{F1}) leads to Eq.\
(\ref{reac-tran-eq-ld}) only for low particle densities. As we shall
demonstrate in Section \ref{simul}, this restriction to low densities severely
limits the efficiency of simulations that use Rule I.

We now employ rule II and show that it leads to the standard reaction-transport
equations independent of particle density. We substitute Eq.~(\ref{F2}) in
Eq.~(\ref{reac-also}) and calculate the expectation value of $F_r(\{{\cal T}
N_\beta(x,t): s_\beta \in {\cal S} \})$. If we assume, as before, molecular
chaos, the expectation value of the product over $\beta$ factorises. If we
further invoke the smoothness assumption and the Poisson distribution which
this
implies ( Eq.~(\ref{prob-tran-n}) ), then the individual terms of the product
have the form
\begin{eqnarray}
< \prod_{m=1}^{\nu_{\alpha r}^{(i)}} ( {\cal T}N_\alpha(x,t) - m + 1 ) > & = &
\sum_{n=0}^{\infty} \prod_{m=1}^{\nu_{\alpha r}^{(i)}}(n-m+1)
\wp ( {\cal T} N_\alpha(x,t) = n ) \nonumber \\
& = & \sum_{n=\nu_{\alpha r}^{(i)}}^{\infty} \prod_{m=1}^{\nu_{\alpha r}^{(i)}}
(n-m+1) \frac{\rho_\alpha^{n}(x,t+\tau/2)}{n!}
e^{-\rho_\alpha(x,t+\tau/2)}
\nonumber \\
& = & \sum_{n=\nu_{\alpha r}^{(i)}}^{\infty}
\frac{\rho_\alpha^{n}(x,t+\tau/2)}
{ \left( n-\nu_{\alpha r}^{(i)} \right) !} e^{-\rho_\alpha(x,t+\tau/2)}
\nonumber \\
& = & \rho_\alpha^{\nu_{\alpha r}^{(i)}} \left( x,t+{\tau \over 2} \right)
e^{-\rho_\alpha(x,t+\tau/2)}
\sum_{n=\nu_{\alpha r}^{(i)}}^{\infty}
\frac{\rho_\alpha^{n-\nu_{\alpha r}^{(i)}}(x,t+\tau/2)}
{ \left( n-\nu_{\alpha r}^{(i)} \right) !}           \nonumber \\
& = & \rho_\alpha^{\nu_{\alpha r}^{(i)}} \left( x,t+{\tau \over 2} \right)
e^{-\rho_\alpha(x,t+\tau/2)}
\sum_{n=0}^{\infty}
\frac{\rho_\alpha^{n}(x,t+\tau/2)}{n!} \nonumber \\
 & = & \rho_\alpha^{\nu_{\alpha r}^{(i)}} \left( x,t+{\tau \over 2} \right) .
\end{eqnarray}
On the RHS of the first equality above, the summation over $n$ need only be
carried out from $n = \nu_{\alpha r}^{(i)}$, because the product over $m$
always vanishes for $0 \leq n \leq \nu_{\alpha r}^{(i)} - 1$.
Repeating the steps of our earlier derivation of the continuum limit, we arrive
at ( cf.\ Eq.\ (\ref{reac-tran-eq1}) )
\begin{equation}
\frac{\partial\rho_\alpha(x,t)}{\partial t} =
- V_\alpha \frac{\partial\rho_\alpha(x,t)}{\partial x}
+ D_\alpha \frac{\partial^2\rho_\alpha(x,t)}{\partial x^2} +
\sum_{r=1}^{R} \left( \nu_{\alpha r}^{(f)} - \nu_{\alpha r}^{(i)} \right) k_r
\prod_\beta \rho_\beta^{\nu_{\beta r}^{(i)}}(x,t)
\hspace{.5cm} \mbox{(Rule II)}, \label{reac-tran-eq2}
\end{equation}
where
\begin{equation}
k_r = P_r / \tau \hspace{1cm} \mbox{(Rule II)}.
\end{equation}
Eq.\ (\ref{reac-tran-eq2}) holds as a strict equality \underline{without} the
additional assumption of low density.

If $\rho_\alpha$ were solute concentrations, this would be the form commonly
used in modelling systems with transport and chemical reactions. We have
already
remarked that particle densities are not the same as measurable concentrations.
Clearly, we do not expect in the foreseeable future to be able to treat
particle
numbers comparable to Avogadro's number. We therefore \underline{assume} that
it
is legitimate to work with numbers of particles that are by orders of magnitude
smaller than Avogadro's number and obtain the respective concentrations
$C_\alpha(x,t)$ by rescaling the densities $\rho_\alpha(x,t)$:
\begin{equation}
C_\alpha(x,t) \equiv \gamma \rho_\alpha(x,t).           \label{conc-def}
\end{equation}
We further assume that the scaling factor $\gamma$ is universal (i.e.\
independent of species, space, time and the value of the density itself). Thus,
we typically fix the value of $\gamma$ from the initial conditions of the
simulation (total particle density) and the real problem it purports to model
(total concentration). We then perform the simulation and recover the
concentrations at the desired times and locations by multiplying the densities
at those times and locations by $\gamma$. If we substitute $\rho_\alpha =
C_\alpha/\gamma$ into Eq.~(\ref{reac-tran-eq2}), we can absorb $\gamma$ in the
rate constants by defining the new rate constants $K_r$:
\begin{equation}
K_r \equiv k_r \gamma^{ - \sum_\alpha \nu_{\alpha r}^{(i)}+1} . \label{kr2}
\end{equation}
Then, the concentrations satisfy equations of the standard form:
\begin{equation}
\frac{\partial C_\alpha(x,t)}{\partial t} =
- V_\alpha \frac{\partial C_\alpha(x,t)}{\partial x}
+ D_\alpha \frac{\partial^2 C_\alpha(x,t)}{\partial x^2} +
\sum_{r=1}^{R} \left( \nu_{\alpha r}^{(f)} - \nu_{\alpha r}^{(i)} \right) K_r
\prod_\beta C_\beta^{\nu_{\beta r}^{(i)}}(x,t).
\label{reac-tran-eqC}
\end{equation}
Here the parameters $V_\alpha, D_\alpha$ and $K_r$ model directly properties of
the physical system. In particular, $K_r$ is identified with the physical rate
constants in Eq.~(\ref{typ-rea}). We note that, if we substitute $\rho_\alpha =
C_\alpha / \gamma$ into Eq.~(\ref{reac-tran-eq1}), then $\gamma$ cancels out if
only linear terms (transport terms and, possibly, linear reactions) are
present,
but in general non-linear terms result in a non-trivial $\gamma$-dependence.
For
low densities, rule I yields Eq.~(\ref{reac-tran-eqC}) in an approximate form,
following from Eq.~(\ref{reac-tran-eq-ld}).

It should be mentioned that even rule II is not completely free of limitations
on the density. The reason is quite different this time and becomes obvious if
one substitutes (\ref{F2}) in Eq.~(\ref{reac-prob}). Since the probability for
$\eta_{x,r} = 1$ must be $\leq 1$ (we refer to this condition as {\em
probability conservation}), the product of $P_r$ and the appropriate
combination
of occupation numbers at any site should not exceed one. For example, if $r$ is
the reaction $a+2b \rightarrow c$, we should have $P_r N_a N_b(N_b-1) \leq 1$.
One  has to make sure, therefore, that particle density is low enough to
guarantee  that the probability for the above product to exceed one is
negligibly small.  Alternatively, since it is the ratio $P_r/\tau$ that is
related to physical  parameters, we can handle arbitrarily high particle
densities by making $P_r$  and, consequently, $\tau$ appropriately small. The
limit in this case is set by  the resulting increase in the number of
iterations
and the accordingly greater  computation time.

In Section \ref{simul} we shall study in detail a reaction-diffusion system in
which particles of the species $a$, $b$ and $c$ interact via the reversible
reaction
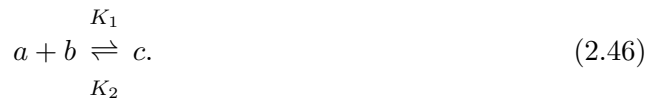
\begin{equation}
a + b \begin{array}{c} \mbox{\scriptsize $K_1$} \\ \rightleftharpoons \\
\mbox{\scriptsize $K_2$} \end{array} c.
\label{ab-c} \end{equation}
The reaction-transport equation for the concentration of species $a$ is
obtained
as a special case of Eq.~(\ref{reac-tran-eqC}):
\begin{equation}
\frac{\partial C_a(x,t)}{\partial t} =
- V_a \frac{\partial C_a(x,t)}{\partial x}
+ D_a \frac{\partial^2 C_a(x,t)}{\partial x^2}
- K_1 C_a(x,t) C_b(x,t) + K_2 C_c(x,t).          \label{reac-tran-exa}
\end{equation}
According to Eqs.~(\ref{kr1}) and (\ref{kr2}), if there is at least one $a$-
and one $b$-particle at a site, the reaction $a+b \rightarrow c$ will occur
with
probability $P_1 = \tau \gamma K_1$. Similarly, if there is at least one
$c$-particle, it will disintegrate into an $a$ and a $b$ with probability $P_2
=
\tau K_2$. If we are in a diffusion-limited regime, where chemical equilibrium
is attained on a much shorter time scale than that of diffusion, then a dynamic
equilibrium is established locally between the reactions $a+b \rightarrow c$
and
$c \rightarrow a+b$, the two rates cancelling each other in
Eq.~(\ref{reac-tran-exa}):
\begin{equation}
K_1 C_a C_b = K_2 C_c \rightarrow \frac{C_c}{C_a C_b} = \frac{K_1}{K_2}
\rightarrow \frac{\rho_c}{\rho_a \rho_b} = \frac{k_1}{k_2} \hspace{0.25cm} ,
\label{mass-act1}
\end{equation}
where we have omitted for simplicity the space and time arguments.
Eq.~(\ref{mass-act1}) is a special case of the {\em law of mass action} for
ideal solutes, with {\em equilibrium constant} ${\cal K} = K_1/K_2$.\
\footnote{The law of mass action in this simple form holds only for infinite
dilution. For higher concentrations, interactions among the species (e.g.\ of
electrostatic nature) complicate the situation. Such effects may in principle
be incorporated in our model, but are not considered in its present form.}

If we use rule I, the exact reaction-transport equations are obtained as a
special case of (\ref{reac-tran-eq1}), for example,
\begin{equation}
{\partial \rho_a(x,t) \over \partial t} =
- V_a {\partial \rho_a(x,t) \over \partial x}
+ D_a {\partial^2 \rho_\a(x,t) \over \partial x^2}
- k_1 \left( 1-e^{-\rho_a(x,t)} \right) \left( 1-e^{-\rho_b(x,t)} \right)
+ k_2 \left( 1-e^{-\rho_c(x,t)} \right),
\label{rate-exa1}
\end{equation}
where $k_1 = \gamma K_1$ and $k_2 = K_2$. We note that no low density
assumption
has been made here. According to (\ref{rate-exa1}), the law of mass action
(Eq.~(\ref{mass-act1})) is replaced by
\begin{equation}
{ 1-e^{-\rho_c} \over
\left( 1-e^{-\rho_a} \right) \left( 1 - e^{-\rho_b} \right) }
= {k_1 \over k_2} \hspace{0.25cm} .
\label{mass-act2}
\end{equation}
For $\rho_a$, $\rho_b$, $\rho_c \ll 1$ we recover the familiar law of mass
action and, if we write $C_a(x,t) \equiv \gamma \rho_a(x,t)$ etc., Eq.\
(\ref{rate-exa1}) reduces to (\ref{reac-tran-exa}) as an approximate equation;
if we are sufficiently close to chemical equilibrium, $K_1 C_a C_b = \gamma k_1
\rho_a \rho_b$ is of the same order of magnitude as $K_2 C_c = \gamma k_2
\rho_c
$ and  we can say that Eq.\ (\ref{reac-tran-exa}) holds to $O(C_c^2/\gamma^2) =
O(\rho_c^2)$.

\subsection{Homogeneous System}   \label{homo}
\indent

In a homogeneous system, the particle density is independent of the spatial
variable and we can write it as $\rho_{\alpha}(t)$. We define
$\rho_{\alpha}(t)$
as the average number of $\alpha$-particles per site:
\begin{equation}
\rho_{\alpha}(t) \equiv { \sum_x N_{\alpha}(x,t) \over {\cal V} }
\hspace{0.25cm} ,
\end{equation}
where $\sum_x N_{\alpha}(x,t)$ is the total number of particles of species
$\alpha$ and $\cal V$ is the total number of sites. In two dimensions, with
$N_x$ sites in the $x$-direction and $N_y$ sites in the $y$-direction,
${\cal V} \equiv N_x \times N_y$. For a homogeneous system, this definition is
equivalent to the ensemble  average used in Section \ref{model}.

We repeat the steps that led from Eq.~(\ref{more-tran-reac-op}) to
Eqs.~(\ref{reac-tran-eq1}) and (\ref{reac-tran-eq2}), but with $< \ldots >$
understood this time as the {\em average over all sites}. The first term on the
RHS of Eq.~(\ref{more-tran-reac-op}) becomes $< {\cal T}N_{\alpha}(x,y,t) > =
\rho_{\alpha}(t) = < N_{\alpha}(x,t) >$ upon averaging, as if the transport
operator $\cal T$ had not been applied at all to the occupation number. This is
natural, as transport does not change the average properties of a homogeneous
system. We factorise the expectation value of $F_r$ in Eq.~(\ref{reac-also})
under the  assumption of molecular chaos (see the discussion following
Eq.~(\ref{reac-also})) and evaluate the individual terms using the Poisson
distribution (derived on the basis of the arguments that led to
Eq.~(\ref{prob-tran-n-also}), but applied this time to the system as a whole).
We thus arrive at an equation for $\rho_{\alpha}(t)$:
\begin{eqnarray}
\addtocounter{zaehler}{1}
{d \rho_{\alpha}(t) \over dt} =
\sum_{r=1}^{R} \left( \nu_{\alpha r}^{(f)} - \nu_{\alpha r}^{(i)} \right) k_r
\prod_{\beta} \sum_{n=\nu_{\beta r}^{(i)}}^{\infty}
{\nu_{\beta r}^{(i)}! \over n!} \rho_{\beta}^n(t) e^{-\rho_{\alpha}(t)}
\hspace{1cm} \mbox{(Rule I)},      \label{rate-eq1}  \\
\addtocounter{equation}{-1} \addtocounter{zaehler}{1}
{d \rho_{\alpha}(t) \over dt} =
\sum_{r=1}^{R} \left( \nu_{\alpha r}^{(f)} - \nu_{\alpha r}^{(i)} \right) k_r
\prod_\beta \rho_\beta^{\nu_{\beta r}^{(i)}}(t)
\hspace{1cm} \mbox{(Rule II)}.     \label{rate-eq2}
\end{eqnarray}
\setcounter{zaehler}{0}

\subsection{Boundary Conditions}
\indent

We shall discuss here the transport behaviour of particles when they reach the
lattice boundary. Since chemical reactions will not enter in the present
discussion, we shall make use only of Eq.\ (\ref{tran-only}). Each boundary
condition will be determined once we formulate the microscopic transport rule
which replaces Eq.\ (\ref{tran-op}) for sites at the lattice boundary.

We first consider an {\em impermeable boundary}, with particles bouncing  back
when they reach it. In one dimension we can set the  boundary at the first site
on the left (say, $x=0$) and the last on the right  ($x = x_{max}$). We
consider
the left end of the lattice. The first two sites  are $x = 0$ and $x = \lambda$
respectively. We define the transport operation  as before by the rule:
$\mbox{\boldmath $\xi$}_{x,n} = (1,0,0), (0,1,0)$ or $(0,0,1)$  with
probability
$p, 1-p-q$ and $q$ respectively, unless $x=0$, in which case $\mbox{\boldmath
$\xi$}_{x,n} = (1,0,0)$ or $(0,1,0)$ with probability $p$ and $1-p$
respectively
($p$ and $q$ are the same as in the interior of the lattice). In other words,
once a particle reaches the left boundary, it moves to the right with the same
probability $p$ as in the interior of the lattice and remains at  the boundary
with probability $1-p$. Equivalently, one may think of the boundary as lying at
$-\lambda/2$ with particles at $x=0$ moving according to the  same rule as in
the interior, but bouncing back to $x=0$ \underline{within the  same transport
step} if they hit the boundary. The  equivalent of Eq.\ (\ref{tran-av}) for
$x=0$ is obtained from the above  probabilities and the fact that there are no
particles coming from the left:
\begin{equation}
\rho(0,t+\tau) \equiv <{\cal T} N(0,t) > = (1-p)\rho(0,t)+q\rho(\lambda,t),
\label{I-0}
\end{equation}
Eq.\ (\ref{I-0}) can be rearranged as follows
\begin{eqnarray}
{\rho(0,t+\tau) - \rho(0,t) \over \tau} & = & {q\rho(\lambda,t) - p\rho(0,t)
 \over \tau} \nonumber \\
 & = & {\lambda \over \tau} \, \left\{(p+q) \;
 {\rho(\lambda,t) - \rho(0,t) \over 2\lambda} - {p-q \over 2\lambda} \;
 \left[ \rho(\lambda,t) + \rho(0,t) \right] \right\} \hspace{0.25cm} .
 \nonumber \\
\mbox{} \label{I-1}
\end{eqnarray}
In the limit $\lambda, \tau \rightarrow 0$, with $p-q \sim O(\lambda)$  and
$\lambda^2/\tau \sim O(1)$, the LHS of Eq.\ (\ref{I-1}) remains finite while on
the RHS $\lambda/\tau \rightarrow \infty$. This implies that the expression
in the curly brackets on the RHS vanishes so that
\begin{equation}
{\rho(\lambda,t) - \rho(0,t) \over \lambda} = {p-q \over (p+q)\lambda} \;
\left[ \rho(\lambda,t) + \rho(0,t) \right] =
{V \over 2D} \left[ \rho(\lambda,t) + \rho(0,t) \right] ,
\end{equation}
where we have used the definitions (\ref{VD}). In the continuum limit we obtain
\begin{equation}
\left. \left[ V \rho(x,t) - D {\partial \rho(x,t)
 \over \partial x} \right] \right|_{x=0} = 0 \hspace{0.25cm} . \label{noflux}
\end{equation}
Eq.\ (\ref{noflux}) is a statement of
the condition that the {\em total flux} of solute (i.e.\ the sum of the {\em
advective flux} $V\rho(x,t)$ and the {\em diffusive/dispersive flux}
$-D\partial\rho(x,t)/\partial x$ ) vanishes at $x = 0$. This is intuitively
clear from the definition of the impermeable boundary: any particles that reach
it bounce off so that  there is no flux across the boundary.

It is common in solute transport problems to specify either the concentration
or
its gradient at the boundary. These or mixed boundary conditions ( i.e.\
relating the concentration with its gradient, such as Eq.\ (\ref{noflux}) ) can
be easily implemented in our simulations once their physical background is
clear. Thus, in the case when the concentration at the $x=0$ boundary is fixed,
it may be assumed that there is a large homogeneous reservoir, extending beyond
the system of interest  and having the given concentration. Introducing
explicitly a reservoir beyond  the boundary at $x=0$ would be correct but
impractical. This situation can be  more efficiently simulated by assigning to
$x=0$ at each transport step an  occupation number from a set of random numbers
obeying the appropriate Poisson  distribution (instead of applying the
simulation rule at $x=0$). A special case is that of a sink, i.e.\ vanishing
concentration, at $x = 0$. The occupation number is then set at all times equal
to zero at $x = 0$ and the evolution rule is applied normally to all sites of
the lattice interior.

Alternatively, one may specify $\left. \partial\rho(x,t) / \partial x
\right|_{x=0} = 0$ as the boundary condition. To simulate this, we add formally
site $x = -\lambda$ to the lattice. Before each transport step we set the
occupation number at $x = -\lambda$ by $N(-\lambda,t) = N(\lambda,t)$. The
evolution rule is then applied to all other lattice sites, including $x = 0$:
\begin{eqnarray}
\rho(0,t+\tau) & = & p\rho(-\lambda,t) + (1-p-q)\rho(0,t) + q\rho(\lambda,t)
\nonumber \\   & = & p\rho(\lambda,t) + (1-p-q)\rho(0,t) + q\rho(\lambda,t)
\end{eqnarray}
which can be rearranged as
\begin{equation}
{\rho(0,t+\tau) - \rho(0,t) \over \tau} =
(p+q) {\rho(\lambda,t) - \rho(0,t) \over \tau} .   \label{I-2}
\end{equation}
Arguing as before, we must have $\rho(\lambda,t) - \rho(0,t) \sim
O(\lambda^2)$,
if the RHS of (\ref{I-2}) is to remain finite in the continuum limit. It
follows
that $\left. \partial \rho(x,t) / \partial x \right|_{x=0} = \left[
\rho(\lambda,t) - \rho(0,t) \right] / \lambda + O(\lambda) \rightarrow 0$,
as $\lambda \rightarrow 0$. In physical terms, we make the diffusive flux
vanish at the boundary by superposing equal and opposite amounts of outgoing
and incoming diffusive flux; this leaves only advection to take care of net
solute transport across the boundary.

A {\em periodic boundary condition} is often computationally convenient and is
used when the precise behaviour of the boundary layer is not important, e.g.\
with translationally invariant homogeneous systems or when the boundary is too
far away to influence the region of interest. We impose a periodic boundary
condition by connecting the two ends of the  lattice, so that particles
crossing
the  right boundary appear automatically on the left boundary and particles
crossing the left boundary appear on the right boundary.

\newpage
\section{Simulation of reaction-transport processes} \label{simul}
\setcounter{equation}{0} \indent

The model formulated in Section \ref{model} is very general and can describe a
wide variety of coupled transport-chemical reaction processes. In this section
our primary aim is to demonstrate the versatility of the model by using it to
simulate various systems. After demonstrating a simple system undergoing both
diffusion and advection, we concentrate on systems of diffusing particles
subject to various chemical reaction schemes. A thorough discussion is given of
the reactions $a + b \rightleftharpoons c$ and $a + b \rightarrow c$; these
reactions are particularly suitable for displaying the essential microscopic
aspects of the simulation. In each case, the results of the simulations are
compared with the ones obtained from the corresponding PDE's~\footnote{The
differential  equations are solved using standard finite difference methods. We
do not intend  to discuss here the convergence of the standard methods. All
results obtained  with them have been checked for convergence and are
indistinguishable for our  purposes from the true solutions of the continuum
equations.}. The purpose of the comparison is to test the validity of the
smoothness and molecular chaos assumptions used to derive the PDE's in Section
\ref{model}, as well as to study how the simulation converges to the continuum
result. Lastly, we show that our model is also capable of simulating complex
autocatalytic reaction-diffusion systems which, under non-equilibrium
conditions, display remarkable spatial and temporal structures.

\subsection{Diffusion and Advection} \label{D-A}
\indent

We first discuss solute transport without chemical reactions. Advection and
diffusion arise as the macroscopic result of a random walk. In fact, from a
microscopic point of view, there is no fundamental difference between the two
processes. If the random walk is unbiased (equal probability of motion in all
directions) there is only diffusion and no advection, whereas a bias in favour
of a certain direction produces diffusion coupled with advection in the chosen
direction. Fig.~\ref{fig1} shows the result of a simulation of solute transport
on a one-dimensional lattice. We simulate simultaneously two solutes with
advection velocities $V_1 = 0.5 \, m/y$ and $V_2 = V_1/10$ and diffusion
coefficients $D_1 = 25 \, m^2/y$ and $D_2 = D_1/10$.~\footnote{The second
solute
can be thought of as being retarded by a factor ${\cal R} = 10$ due to sorption
on the surface of the solid matrix. If one assumes instantaneous sorption
equilibrium and a linear relationship between liquid and solid phase
concentrations, the effect of sorption can indeed be reduced to a retention
factor $\cal R$ that divides the transport coefficients $V_{\alpha}$ and
$D_{\alpha}$.} $\;$ Taking $\tau = 10^{-3}\, y$ and $p_1 + q_1 = 1$, we use the
second of Eqs.~(\ref{VD}) to determine $\lambda$. The probabilities $p_1$,
$q_1$, as well as $p_2$, $q_2$, are calculated from Eq.~(\ref{pq}). Initially
both concentrations are equal to 1 (arbitrary units) for $x \leq 0$, and $0$
for
$x > 0$. The simulation begins with 200 particles of each species per site to
the left of $x = 0$ and the concentration is calculated by averaging the
occupation number over cells of 100 lattice sites and normalising by a factor
$\gamma = 1/200$. The boundary condition is uninteresting here, because the
boundary is chosen far enough, so that its influence does not reach the
displayed region by the time considered. In the figure, the solution of the
transport equation (\ref{tran-eq}) is shown for comparison. Error bars of
length equal to one standard  deviation were estimated and were always found to
be smaller than the plotting  symbol. The small fluctuations around the solid
curve are of statistical nature and diminish if we increase the number of
particles used in the simulation.

\begin{figure}[t]
  \vspace{9.0cm}
  \caption{Result of one-dimensional simulation (species 1:\ solid circles,
  species 2:\ open circles) compared with the solution of the transport
equation
  (species 1:\ solid curve, species 2:\ dashed curve) at $t = 240 y$.
Parameters
  and initial conditions are given in the text.
\label{fig1} }
\end{figure}

\subsection{ ${\bf a + b \protect\rightleftharpoons c}$ ,  homogeneous case}
\indent

We introduce chemical reactions by looking first at a system of particles which
is initially homogeneous. We begin the simulation by placing particles on a
two-dimensional lattice according to a uniform random distribution. As we saw
in
Section~\ref{model}, for a homogeneous system the spatial derivatives of the
density vanish and the reaction-transport equations (\ref{reac-tran-eq1}) and
(\ref{reac-tran-eq2}), corresponding to reaction rules I and II respectively,
reduce to the chemical rate equations (\ref{rate-eq1}) and (\ref{rate-eq2}),
which describe the evolution of uniform particle densities. It is important to
note that these equations have been derived under the assumption of molecular
chaos. In reality, correlations between particles and density fluctuations do
occur, and on occasion, notably in some irreversible reactions as well as in
some autocatalytic systems maintained far from equilibrium, lead to
inhomogeneities, even for systems which initially are homogeneous. Later we
shall discuss such reaction schemes; then of course, the average particle
density need not follow Eqs.~(\ref{rate-eq1}) and (\ref{rate-eq2}). For the
moment, however, we consider the case where the system remains homogeneous as a
function of time, and we address the question of how microscopic dynamics
drives
a discrete system towards equilibrium.

\subsubsection*{(a) Reaction rule I}
\indent

We consider a homogeneous system of $a$-, $b$- and $c$-particles reacting via
the reversible reaction $a + b \rightleftharpoons c$. At each time step every
particle moves randomly to one of the four nearest sites. Particles react
according to rule I: if there are at least one $a$- and one $b$-particle at a
site, then, with probability $P_1$, one $a$- and one $b$-particle are removed
and a $c$-particle is added; whereas, if there are one or more $c$-particles at
a site, then one of them  is replaced by an $a$- and a $b$-particle with
probability $P_2$. As long as we discuss homogeneous systems, we shall average
densities over the whole lattice and express them only as functions of time,
$\rho_{\alpha}(t)$ (cf.~Subsection  \ref{homo}). We follow the approach to
equilibrium by looking at the time  evolution of the {\em reaction quotient}
$\rho_c(t)/\rho_a(t)\rho_b(t)$.

Let $p_{\alpha x}$ ($q_{\alpha x}$) be the probability that an
$\alpha$-particle
moves by one lattice spacing $\lambda$ to the right (left) along the
$x$-direction and $p_{\alpha y}$ ($q_{\alpha y}$) the probability that it moves
up (down) by the same distance along the $y$-direction in one transport step.
Assuming all species to have the same diffusion coefficients, we take
$p_{\alpha x} + q_{\alpha x} + p_{\alpha y} + q_{\alpha y} = 1, \, \forall
\alpha$, i.e.\ particles always move to a neighbouring site.

Here, as in all cases below, we consider the case of no advection for
convenience; when necessary, advection can be easily included in our model as
demonstrated in Subsection \ref{D-A} above. Thus we set $V_{\alpha x} \equiv
(p_{\alpha x}-q_{\alpha x})\lambda / \tau = 0$ and $V_{\alpha y} \equiv
(p_{\alpha y} - q_{\alpha y})\lambda /  \tau = 0$, where $V_{\alpha x}$ and
$V_{\alpha y}$ are the advection velocities in the $x$- and $y$-directions
respectively (thus $p_{\alpha x} =  q_{\alpha x}$ and $p_{\alpha  y} =
q_{\alpha y}, \,\forall \alpha$). We further assume equal diffusion
coefficients
in the $x$- and  $y$-directions ($D_{\alpha x} \equiv  (p_{\alpha x} +
q_{\alpha
x}) \lambda^2 / 2\tau = D_{\alpha y} \equiv  (p_{\alpha y} + q_{\alpha y})
\lambda^2 / 2\tau$).  Putting together the constraints on the various
displacement probabilities, we  deduce $p_{\alpha x} =  q_{\alpha x} =
p_{\alpha
y} = q_{\alpha y} = 1/4$ and  $D_{\alpha x} =  D_{\alpha y} = \lambda^2 / 4\tau
\equiv D$. For convenience, these conditions will hold for all two-dimensional
systems in this paper. We begin the simulation with $50 \, 000$ particles of
each species on a two-dimensional lattice of $500 \times 500$ sites (i.e.\ a
density of 0.2 particles of each species per site) with periodic boundary
conditions. The reaction probabilities are determined through the relations
$k_r = P_r / \tau$, where the rate constants are taken to be $k_1 = 0.8$ and
$k_2 = 0.2$.

For our initial simulation we take $\tau=1$ (arbitrary units) and examine the
time development of the reaction-diffusion system up to time $t=20$. In
Fig.~\ref{fig2a} the solid circles denote the mean value of the reaction
quotient for an ensemble of $21$ systems. The estimated error bars are smaller
than the plotting symbol. The solid curve is obtained by solving the rate
equations (\ref{rate-eq1}). According to the latter, the reaction quotient
reaches the equilibrium value $3.50$ at roughly $t = 10$ and remains constant
thereafter. This value can also be derived by solving Eq.~(\ref{mass-act2})
subject to the constraints $\rho_a(t) - \rho_b(t) = constant$ and $\rho_a(t) +
\rho_c(t) = constant$, which obviously hold for $a + b \rightleftharpoons c$.
The result of the simulations evolves instead towards a slightly higher
equilibrium value and then runs parallel to the solid curve. {\em The
simulations lead to an equilibrium state with relatively more $c$-particles
than
predicted by the rate equations}.

\addtocounter{zaehler}{1}
\begin{figure}[tp]
  \vspace{9.0cm}
  \caption{Time dependence of the reaction quotient for a two-dimensional,
  homogeneous system of $a$-, $b$- and $c$-particles, reacting via $a+b
  \protect\rightleftharpoons c$ according to rule I: mean value over $21$
  simulations with $\tau=1$
  (solid circles), exact result of differential rate equations (solid curve)
and
  approximate result of FDE's (dashed curve). Also shown is the result of $21$
  simulations with $\tau = 1/5$ (open circles). The values of the transport
  and reaction parameters, as well as the initial conditions, are given in the
  text. \label{fig2a} }
\addtocounter{figure}{-1} \addtocounter{zaehler}{1}
  \vspace{9.0cm}
  \caption{The same as Fig.\ \protect\ref{fig2a}, but with $10$ d.p.r.
  \label{fig2b} }
\end{figure}
\setcounter{zaehler}{0}

That the discrete result does not agree with the exact solution of the rate
equations is due to two reasons. As far as the rate equation is concerned, the
problem is completely specified by the initial conditions, the rate constants
$k_r$, and the final time of interest, $t = 20$. Apparently, the time
discretisation used in the simulation is too coarse to provide agreement with
these rate equations. To illustrate this, we solve the coupled FDE's obtained
from Eqs.~(\ref{rate-eq1}) with the same time step, $\tau = 1$, as used in the
simulations~\footnote{The FDE's are obtained by substituting
$\left[ \rho_{\alpha} (t+\tau) - \rho_{\alpha}(t) \right] / \tau$ for
$d\rho_{\alpha}(t)/dt$ in (\ref{rate-eq1}).}. The result of the FDE's (dashed
line in Fig.~\ref{fig2a}) lies far from the exact result it is approximating
(showing that the time discretisation is not sufficient); but it also lies far
from the result of the simulations (although the latter use the same time
discretisation as the FDE's).

The discrepancy between the simulations and the solution of the FDE's is of a
different origin and can be understood as follows: The rate equations and the
approximating FDE's rely on the assumption that $a$- and $b$-particles are
uncorrelated. In the simulations, this assumption does not hold. At the rate of
one diffusion step per reaction step (d.p.r.), the $a$- and $b$-particles
originating from the disintegration of a $c$-particle remain sufficiently close
to each other for the probability of their meeting again (with a subsequent
chance of reaction) to be significantly higher than in a uniform distribution
of
particles, i.e.\ if the assumption of molecular chaos were valid. The reverse
reaction, $c \rightarrow a + b$, is not affected by correlations. As a result,
$a + b \rightarrow c$ is enhanced with respect to $c \rightarrow a + b$ and
relatively more $c$-particles are produced, resulting in a higher reaction
quotient. The conditions of molecular chaos can be systematically approached by
performing an increasing number, $n_D$, of d.p.r. In each evolution step of our
simulation, we repeat the transport operation $n_D$ times before we perform the
reaction operation once. This multiplies the diffusion coefficient by a factor
$n_D$ in the continuum limit, but, for a homogeneous system, the rate equations
remain the same. The reaction quotient in the asymptotic state is shown in
Fig.~\ref{fig3} for $n_D = 1,2,3,5$ and 10. To obtain the value for $n_D =
\infty$, we replace the transport operation by a random redistribution of
particles \footnote{We have tested the validity of this identification by
letting a system of particles diffuse from different initial distributions.
After sufficient iterations of our diffusion algorithm, the occupation number
obeys essentially the Poisson distribution over the accessible lattice sites,
as expected from a uniform random distribution of particles.}. We see that,
{\em as increased diffusion washes out correlations between the particles, the
equilibrium reaction quotient approaches the value obtained from the rate
equations}. Similar conclusions concerning correlation effects have been drawn
in Ref.~\cite{db}.

\begin{figure}[htp]
  \vspace{9.0cm}
  \caption{Reaction quotient for the system of Fig.\ \protect\ref{fig2a},
  averaged over 9000 time steps in the steady state of a simulation,
  for different numbers of diffusion steps per reaction step.
  \label{fig3} }
\end{figure}

The time evolution of the reaction quotient for $n_D = 10$ is shown in
Fig.~\ref{fig2b} (solid circles). As expected from the preceding paragraph,
correlations play a much lesser r\^ole and the system reaches a steady state
only slightly higher than that predicted by the rate equations. However, for
times $t \leq 5$, there is a serious discrepancy with the exact
solution of the rate equations. Comparing again with the solution of the FDE's
(dashed  curve), we find good agreement, thus confirming the smallness of
residual  correlations. In fact, the agreement becomes perfect if we
redistribute  particles randomly between reaction steps ($n_D = \infty$).

We now look at the way our discrete model approaches the continuum limit.
As we refine the time step, we expect to obtain the exact prediction of the
rate
equations, in very much the way the solution of an FDE converges to the exact
solution. We perform simulations with a shorter time step $\tau = 1/5$.
Then we also have to scale down the reaction probabilities $P_1$ and $P_2$ by a
factor 5, in order to obtain the same rate constants $k_1$ and $k_2$. The
results of the simulations executed with these parameters are shown as open
circles in Fig.\ \ref{fig2a} ($n_D = 1$) and in Fig.\ \ref{fig2b} ($n_D = 10$).
The reaction quotient is here averaged over $21$ independent simulations and
statistical errors are smaller than the plotting symbol.

The finer discretisation obviously leads to better agreement with the exact
result in both Figs.\ \ref{fig2a} and \ref{fig2b}. In the case of Fig.\
\ref{fig2b}, where correlations are almost absent, the improvement over the
previous set of simulations (solid circles) is attributable to the use of a
smaller time step, in the sense of convergent FDE schemes. In the case of
Fig.\ \ref{fig2a}, however, the finer discretisation improves the agreement
also
partly by reducing indirectly the effect of correlations. This can be seen as
follows: As the reaction probabilities $P_r$ decrease (by the same factor, say
$m$, as the time step), particles react more weakly in a single  update. This
is
compensated by exactly that many more updates within a given time span, so that
the overall reaction strength (rate constant) remains intact. The particles
execute, however, $m$ times more movements to neighbouring sites in a given
time
span. As a result of this increased agility, particles originating at the same
site are more inclined to lose memory of their common origin and correlations
are subsequently weakened. In fact, {\em as the continuum limit is approached,
any correlations between the $a$- and $b$-particles will disappear}.

Using reaction rule I (Eq.~(\ref{rate-eq1})), we obtain a reaction quotient of
3.50 at chemical equilibrium, for the particular parameters chosen. The
standard
rate law, Eq.~(\ref{rate-eq2}), leads to a significantly different value of
$k_1/k_2 = 4.00$. This difference is obtained for an initial particle density
of
0.2. It is clear that the particle density has to be significantly lower than
0.2, if we intend to reproduce the standard law by using Rule I. In fact a
density of 0.005 would  be necessary to reduce the discrepancy at equilibrium
below $1 \%$. For low particle densities the statistics is poor for each
individual simulation and one has to perform many of them or use a bigger
lattice. On the other hand, one can also obtain better statistics if higher
particle densities can be used. We therefore proceed to investigate reaction
rule II, which allows us to use higher densities.

\addtocounter{zaehler}{1}
\begin{figure}[tp]
  \vspace{9.0cm}
  \caption{Time dependence of the reaction quotient for a two-dimensional,
  homogeneous system of $a$-, $b$- and $c$-particles, reacting via $a+b
  \protect\rightleftharpoons c$ according to rule II: mean value over $21$
  simulations (open circles) and exact result of differential rate equations
  (solid curve). \label{fig4a} }
\addtocounter{figure}{-1} \addtocounter{zaehler}{1}
  \vspace{9.0cm}
  \caption{The same as Fig.\ \protect\ref{fig4a}, but with an initial density
of
  1 and appropriate rescaling of parameters and densities as explained in the
  text. \label{fig4b} }
\end{figure}
\setcounter{zaehler}{0}

\subsubsection*{(b) Reaction rule II}
\indent

We first apply rule II to a system of particles with the same initial density,
0.2, as with rule I. Fig.~\ref{fig4a} shows the result of 21 simulations, on a
$500 \times 500$ lattice, of a homogeneous system of $a$-, $b$- and
$c$-particles diffusing with $n_D = 1$ and reacting via $a + b
\rightleftharpoons c$, according to rule II. Transport and kinetic parameters,
as well as initial and boundary conditions, are identical with those of the
previous simulations on such a lattice. The time step is taken to be $\tau =
1/ 10$ and the reaction parameters $P_1 = 0.08$ and $P_2 = 0.02$.
According to rule II, if there are $n_a$ $a$-particles and $n_b$ $b$-particles
at a site, then one $a$- and one $b$-particle are removed and a $c$-particle
is added with probability $P_1 n_a n_b$, whereas, if there are $n_c$
$c$-particles, one of them is replaced by an $a$- and a $b$-particle with
probability $P_2 n_c$. A smaller time step is dictated by the need to make
$P_1$ and $P_2$ sufficiently small so that the
probability conservation conditions ( cf.\ Eqs.~(\ref{reac-prob}) and
(\ref{F2}) ) \begin{eqnarray}
\addtocounter{zaehler}{1}
 & &\wp (\eta_{x,1}=1 | \{ N_{\alpha}(x,t) = n_{\alpha} \} ) = P_1 n_a n_b
\leq 1
\label{prob-exa0} \\
\addtocounter{equation}{-1} \addtocounter{zaehler}{1}
& {\rm and} &\wp (\eta_{x,2}=1 | \{ N_{\alpha}(x,t) = n_{\alpha} \} ) =
P_2 n_c \leq 1.
\end{eqnarray}
\setcounter{zaehler}{0}
will be respected in all but an insignificant number of cases.
For $P_1 = 0.08$, condition (\ref{prob-exa0}) is violated if $n_a n_b > 12$.
Assuming a Poisson distribution for the occupation numbers, we readily estimate
that $n_a n_b > 12$ occurs at an arbitrary site with probability $0.8 \times
10^{-8}$~\cite{kb}. From this we also estimate that there is a $0.2 \%$
likelihood that probability conservation will be violated at all (i.e.\
anywhere on the lattice) in one iteration step. This likelihood is negligible
for practical simulations and, in the rare cases when the probability of
reaction does exceed one, we can safely set this probability equal to one. Due
to the substantially smaller time spacing ($\tau=1/10$), correlation effects
are
insignificant even for 1 d.p.r.\ in this case.

Moving to a higher particle density, we simulate, on a $200 \times 200$
lattice,
a homogeneous system of $a$-, $b$- and $c$-particles, reacting via $a + b
\rightleftharpoons c$, with an initial density of 1 particle of each species
per
site and periodic boundary conditions. The time step is $\tau = 1/10$, as
in the last set of simulations, and we perform 1 d.p.r. To obtain the same rate
equations and initial conditions as before (and hence the same continuum result
to compare with), we have to first rescale the previous value of $P_1$ by a
factor $\gamma = 0.2$ (leaving $P_2$ intact), then perform the simulation and
finally multiply the resulting densities by the same factor. The scaling factor
$\gamma$ is a dimensionless version of the factor used to relate particle
density to concentration in Section \ref{model}. The result of $21$ simulations
is shown in Fig.~\ref{fig4b}. In this case, probability conservation is
violated
if $n_a n_b > 62$, which occurs with probability $0.3 \times 10^{-9}$ at a
given
site and with probability $0.1 \times 10^{-4}$ anywhere on the lattice during
an iteration step.

\subsubsection*{(c) Optimisation of Rule II}
\indent

We have seen that rule I approximates the standard rate law only for
sufficiently low particle densities. Under this constraint, the quality of
statistics can be improved either by performing more simulations, or by
increasing the number of lattice sites. Rule II, however, leads to the standard
rate law without the restriction of low particle densities and, in that case,
better statistics can also be achieved by utilising the higher densities.
On the other hand, rule I is simpler and its simulation computationally more
efficient for a given density. It is very important to assess whether the
additional freedom (wider density range) afforded by rule II can be exploited
in order to obtain higher computational efficiency (for the same statistics).
More  precisely, we address the questions: (i) whether there is an optimal
density which minimises the computation time needed to perform a simulation
using rule II, (ii) how this computation time compares with the time required
by
rule I. We shall not attempt to provide a universal answer to these questions;
instead, we concentrate on the example of $a+b\rightleftharpoons c$ we have
been
considering so far, emphasising the conclusions that have wider validity.

We perform a series of simulations of the above system of $a$-, $b$- and
$c$-particles on a $200 \times 200$ lattice, using rule II and different values
of the initial particle density, which we take to be the same for all species:
$\rho_a(0) = \rho_b(0) = \rho_c(0) \equiv \rho$. The reaction parameters $P_r$
are rescaled each time so that the continuum rate equations and initial
conditions are the same as those in the last set of simulations above. We
finally rescale the densities $\rho_\alpha(t)$ obtained from the simulation by
the ratio of the reference density 0.2 to the initial density of particles on
the lattice, $\rho$, thus $\rho_\alpha(t) \longrightarrow
\tilde{\rho}_\alpha(t) \equiv (0.2 / \rho)  \rho_\alpha(t)$. In each
case we calculate the reaction quotient  $\tilde{\rho}_c(t) / \tilde{\rho}_a(t)
\tilde{\rho}_b(t)$ and compare its asymptotic value with the result obtained by
solving Eq.~(\ref{rate-eq2}) with $k_1 = 0.8$, $k_2 =  0.2$, $\rho = 0.2$ and
the stoichiometric coefficients appropriate for $a + b \rightleftharpoons c$.

We have already seen that the result of the simulations lies arbitrarily close
to that of the rate equations for the right choice of $\tau$. Here we require
that the result of the simulations does not deviate by more than $\sim 10-15
\%$
from the exact result. This is roughly the discrepancy we obtain if we solve
Eq.~(\ref{rate-eq1}) with the same parameters (resulting in a reaction quotient
of 3.50 instead of 4.00, as given by the standard rate law). This discrepancy
reflects the systematic error expected from  rule I, since the densities
calculated with that rule converge to values that satisfy Eq.~(\ref{mass-act2})
instead of the standard Eq.~(\ref{mass-act1}).

\begin{figure}[tp]
  \vspace{9.0cm}
  \caption{(a) Normalised minimum computation time for one simulation, as a
  function of initial particle density: Rule II (crosses, curve to guide the
  eye) and rule I (solid circle). (b) Estimated minimum computation time
  ($T_c = c / \rho P_1$), corresponding to a $15 \%$ likelihood of
  probability-conservation violation at a site, as a function of particle
  density, with $c$ fitted to actual $T_c$ at $\rho = 10$.  \label{fig5} }
\end{figure}

Given the above maximum acceptable error, we seek to minimise the computation
time by reducing the number of iteration steps. This is equivalent to
increasing
$\tau$, which in turn implies a larger $P_1$ (of course, also a larger $P_2$,
but this has no influence on most of our considerations). Increasing $\tau$ can
have a number of possible effects on the results of the simulation. Firstly, we
expect deviation from the rate equation in the sense that the corresponding FDE
deviates from the rate equation. However, it can be easily checked that the FDE
gives the same reaction quotient \underline{at equilibrium} as the rate
equation, independently of the value of $\tau$. Since we have chosen to use
only
this asymptotic ratio for comparison of simulations, increasing $\tau$ will not
give rise to such numerical convergence problems in our case. On the other
hand,
an increase in the probability of reactions leads to stronger correlations
between the particles, and this does affect the reaction quotient at
equilibrium. By varying the number of diffusion steps per reaction step, we
have
actually found that the effect of such correlations is relatively small. We
find
instead that the main effect of increasing $\tau$, and hence $P_1$, is that
probability conservation is violated more frequently. Since an unphysical
probability ($> 1$) is treated in our simulation as if it were 1, the reaction
$a + b \rightarrow c$ is partly suppressed each time $P_1$ exceeds 1.  This
amounts to a suppression of the production of $c$-particles and appears
macroscopically as a reduction of the reaction quotient.

For a given initial density $\rho$, we obtain our minimum computation time,
$t_c(\rho)$, when the reaction quotient is reduced by the maximum amount
allowed, namely by the factor $10-15 \%$ discussed above. For all densities we
normalise this minimum computation time by demanding the same statistics as for
$\rho = 0.2$; thus we introduce the normalised minimum computation time
$T_c(\rho)$ by the equation $T_c(\rho) = \gamma t_c(\rho)$, where $\gamma =
0.2
/ \rho$. In Fig.~\ref{fig5}a we show $T_c$ as a function of the initial density
$\rho$: there is a minimum for $\rho$ between 3 and 5. Thus, {\em there is a
range of densities for which rule II can be simulated with maximum efficiency}.
To compare this with the efficiency of rule I, we also show the computation
time
for a simulation of that rule, performed with initial density 0.2 and the same
number of iteration steps as the optimal simulation of rule II with the same
density. We see that, {\em although rule I may be significantly more efficient
than rule II for the same density, it can be less efficient than rule II when
the latter is used with a higher density}. In other words, using rule~II with
higher densities can offer an advantage in computational efficiency (in the
particular case we are considering, by a factor of 4-5).

In Fig.~\ref{fig5}a we do not continue to higher densities because probability
conservation begins to be violated due to the size of $P_2$ and the picture
becomes more complex beyond $\rho \simeq 10$. We expect however that the
upward trend of the curve in Fig.~\ref{fig5}a continues to higher densities. To
show that higher densities result in ever increasing computation times we argue
as follows: It is plausible to assume that, for high particle densities $\rho$,
the computation time $T_c$ required for given statistics is roughly
proportional
to $N_s \times N_t \times N_p$, where $N_p$ is the total number of particles,
$N_t$ the number of iteration steps in a single simulation and $N_s$ the number
of independent simulations required to achieve the desired statistics. Since
$N_s \propto 1/\rho$ and $N_p \sim \rho N_x^2$ (assuming a rectangular $N_x
\times  N_y$ lattice, with $N_x$ of the same order of magnitude as $N_y$), we
have $T_c  \propto N_x^2 \times N_t$. But $N_x = L/\lambda$ and $N_t = T/\tau$,
where $L$  is the linear size of the system and $T$ the time interval during
which the  system evolves. For a reaction of the type $a + b \rightarrow c$, we
saw below  Eq.~(\ref{reac-tran-exa}) that $\tau = P_1/\gamma K_1$, where
$\gamma$ relates  $\rho$ to the physical concentration $C$: $C = \gamma \rho$
(cf.\  Eq.~(\ref{conc-def})); from Eq.~(\ref{VD}) we also find $\lambda \propto
\sqrt{D\tau}$. Putting things together, we deduce $T_c \propto 1/(\rho P_1)^2$.
Since our aim is to minimise computation time, $P_1$ has to be as high as
possible, with the upper bound set by the maximum acceptable likelihood of
probability-conservation violation. Given the density, we can calculate the
violation probability for any given value of $P_1$ assuming a Poisson
distribution of the occupation numbers. Conversely, if we fix the violation
probability, we can determine the corresponding value of $P_1$.  To be
definite,
let us set the violation probability at a site to be $15 \%$. It is then found
that $P_1$ decreases, as a function of $\rho$, faster than $1/\rho$, which
implies that $1/(\rho P_1)^2$, and hence $T_c$, is an increasing function of
$\rho$. The decrease of $P_1$, and hence of $\tau$ and $\lambda$, with $\rho$
results in an increase of lattice size and iteration number which in turn
implies an increase in computation time. There is a subtle point in the above
argument which deserves mentioning. For a homogeneous system we need not
require
a fixed diffusion coefficient $D$ and hence we can keep $\lambda$ and $N_x$
fixed while we vary $P_1$ and $\tau$. This is in fact what we did in the the
simulations described above. Then $T_c \propto N_t \propto 1 / \rho P_1$, which
is shown in Fig.\ 5b, is also an increasing function of $\rho$. We conclude
that
there is no advantage in increasing the density beyond the minimum in
Fig.~\ref{fig5}a.

\subsection{${\bf a+b \rightarrow c}$, homogeneous case}
\indent

An interesting situation arises when we turn off the reverse reaction. Then the
problem consists of $a$- and $b$-particles that diffuse and combine to form
inert $c$-particles upon meeting. Equivalently, we may neglect species $c$
altogether and think in terms of an annihilation reaction between `particles'
$a$ and `antiparticles' $b$.

We assume the numbers of $a$- and $b$-particles to be initially equal
($\rho_a(0) = \rho_b(0) \equiv \rho$), which implies that they will remain so.
Then, the density $\rho_a(t) = \rho_b(t)$ obeys the rate equation $d\rho_a/dt =
-k \rho_a^2$, where $k \equiv P_1 / \tau$ is the rate constant of the reaction
$a + b \rightarrow c$ (of course $P_2 = 0$). The solution of the rate equation
is $\rho_a(t) = \rho / ( 1 + k\rho t )$, which behaves asymptotically in time
as
$t^{-1}$. It is known, however, that, in microscopic simulations, the long-time
behaviour of the system is determined by long-range density fluctuations, which
give rise to an asymptotic $t^{-1/2}$ behaviour in two dimensions. Following
Ref.~\cite{tw}, this can be understood as follows: Initially ($t = 0$),
particles $a$ and $b$ are distributed randomly with uniform probability. At a
much later time $t$, particles have spread over a length scale $\ell_D \equiv
(Dt)^{1/2}$ due to diffusion and fluctuations on a longer length scale have not
had time to dissolve. In a two-dimensional region of linear size $\ell_D$,
there
are initially on the order of $\rho \ell_D^{2}$ particles of each species;
however, due to statistical fluctuations, there will be deviations from this
number. Assuming variations of one standard deviation, one can expect the
difference between the numbers of $a$- and $b$-particles to be of order $(\rho
\ell_D^2)^{1/2}$. Therefore, in this region of volume $\sim \ell_D^2$, there
is an excess density of $a$- or $b$-particles on the order of $\rho^{1/2} /
\ell_D$. For a given rate constant $k$, we can always choose the time $t$
large
enough so that the bulk of the particles will have annihilated and only  the
excess density will remain by that time. Thus, what remains of the initial
system are regions occupied by one or the other species, with density of order
$\ell_D^{-1} \sim t^{-1/2}$; the reaction takes place only along the boundaries
of these regions and is slower than under the well-mixed conditions assumed by
the rate equation. Thus, {\em the initial fluctuations in particle density give
rise to a decay mechanism which is slower than $t^{-1}$ and dominates at
sufficiently long times. The rate equation, by neglecting fluctuations, does
not
account for this mechanism and predicts the wrong long-time behaviour}.

Simulations using our model reproduce asymptotically this result. At finite
times, we observe that the deviation from the result of the rate equations
depends on the details of the experiment: (i) it decreases when the diffusion
constant increases, since this limits the range of contributing fluctuations,
and (ii) it increases with the rate constant, as the bulk of the particles
annihilate faster and clusters develop sooner (Fig.~\ref{fig6}, where $\tau =
1$).

\begin{figure}[tp]
  \vspace{9.0cm}
  \caption{Average density of a two-dimensional, homogeneous system reacting
  via $a+b \protect\rightarrow c$ according to rule II, as a function of time,
  for different values of the rate constant: simulation (solid curves) compared
  with rate equation (dotted curves).  \label{fig6} }
  \vspace{9.0cm}
  \caption{Width of reaction zone and of distribution of annihilation
  events for a two-dimensional, initially separated system reacting via $a+b
  \protect\rightarrow c$ according to rule II, as a function of time: solution
  of reaction-transport equations (dotted, dashed and dot-dashed curves for 1,
  4 and 10 d.p.r.\ respectively) compared with simulation (solid curves).
  \label{fig7} }
\end{figure}

\subsection{${\bf a + b \protect\rightleftharpoons c}$, at an interface}
\indent

Moving away from homogeneous systems, we consider a system of $a$- and
$b$-particles which initially ($t = 0$) occupy the $x \geq 0$ and $x \leq 0$
halves of a two-dimensional ($N_x \times N_y$) lattice respectively. For
$t > 0$, they diffuse and react, producing $c$-particles ($a + b \rightarrow
c$); the latter diffuse and disintegrate into $a$- and $b$-particles ($c
\rightarrow a + b$). In the process, the $a$- and $b$-particles diffuse into
the
regions of each other, while $c$-particles form a distribution peaked at $x =
0$. Here and in what follows, reaction rule II is used. We perform simulations,
in which there are initially  $50 \, 000$ particles on each half of a $500
\times 500$ lattice (i.e.\ 0.4 particles/site) and other parameters are the
same
as before. Since the macroscopic problem is effectively one-dimensional for the
particular initial condition, we define particle density as a function of time
and the spatial variable $x$ only, by averaging over the occupation number
along
the $y$-direction:
\begin{equation}
\rho_{\alpha}(x,t) \equiv { 1 \over N_y } \sum_{y} N_{\alpha}(x,y,t)
\hspace{0.25cm} .
\end{equation}

We wish to express quantitatively the size of the reaction front, i.e.\ the
region around $x = 0$ where reactions take place. This is obviously related to
the spatial overlap of the densities of reacting particles. As $a$- and
$b$-particles spread on the lattice, their overlap steadily increases
\footnote{We always choose the size of the system so that the overlap does not
reach the boundary during the simulation; the precise boundary condition is,
therefore, irrelevant for the present discussion.}. Using $\rho_{\alpha}(x,t)$,
we define the {\em width of the reaction zone}, $w$, as the standard deviation
of the product of the $a$- and $b$-particle distributions:
\begin{equation}
\bar{x}(t) = {1 \over N_x} \sum_{x} x \rho_a(x,t) \rho_b(x,t) \hspace{1cm} ,
\hspace{1cm} w^2(t) = {1 \over N_x} \sum_{x} \left[ x - \bar{x}(t) \right]^2
\rho_a(x,t) \rho_b(x,t) \hspace{0.25cm} .
\label{width}
\end{equation}
The width obtained from these simulations is consistent, within statistical
fluctuations, with the asymptotic time dependence, namely $t^{1/2}$, predicted
by the reaction-diffusion equations. This behaviour is understood if we note
that our choice of parameter values corresponds to a diffusion-limited regime.
Then, particles essentially spread according to the $t^{1/2}$ law expected from
pure diffusion, with the relative abundance of different species determined
locally by chemical equilibrium. If the reaction probabilities are much smaller
than 1, then the time dependence of $w$ displays more variety at times short on
the time scale of the reactions,  but the asymptotic behaviour remains the
same~\cite{cdkr}.

\subsection{${\bf a + b \protect\rightarrow c}$, at an interface}
\indent

We now turn off the reverse reaction, as in the homogeneous case above, keeping
all other parameters unchanged. In the diffusion-limited regime, $a$-$b$
annihilation around $x = 0$ proceeds so fast that new particles do not have the
time to get there by diffusion. As a result, a {\em depletion zone} develops
around $x = 0$, where the densities of $a$- and $b$-particles are significantly
smaller than their initial values. The width of this zone grows as $t^{1/2}$.
It can be shown from the reaction-diffusion equations that the width of the
reaction zone, as defined previously, varies asymptotically with time as
$t^{1/6}$~\cite{gr}. For reasons that will become clear shortly, we define
alternatively the {\em width of the spatial distribution of annihilation events
up to time t}. Each time an $a$- and a $b$-particle annihilate, we keep a
record
of the event and then we count at each lattice site all events that took place
up to time $t$. The distribution of annihilation events is identical with the
distribution of $c$-particles, provided the latter remain permanently at the
site where they are produced. The new width, $w'$, is then defined by replacing
in (\ref{width}) the product of the $a$- and $b$-densities by the density of
the
inert, immobile $c$-species. It can be easily shown that $w'$ behaves
asymptotically also as $t^{1/6}$ according to the reaction-diffusion equations.

In Fig.~\ref{fig7} we compare the result of the simulation (solid curves) with
that of the reaction-diffusion equations: the dotted curves describe the time
evolution of the widths of the reaction zone ($w$, upper curve) and the
distribution of reaction events ($w'$, lower curve) for 1 d.p.r., while the
dashed and dot-dashed curves describe the same quantities for 4 and 10 d.p.r.\
respectively. The comparison is clearly facilitated by the better statistics in
the case of $w'$. We notice that for 1 d.p.r.\ the result of the simulation
lies
above that of the reaction-diffusion equations; moreover, the former apparently
grows with a higher time exponent than the expected 1/6. This is consistent
with
Ref.~\cite{cd}. It is clear that, with enhanced diffusion (4 and 10 d.p.r.),
the
result of the simulation converges systematically to that of the continuum
equations. Continuing the simulation to greater times modifies slightly the
time
exponents, which depend further on the dimensionality of the lattice~\cite{ccd}
and the reaction strength, but the following qualitative picture remains: {\em
in the absence of sufficient diffusive mixing, $a$ and $b$ annihilate less
strongly and hence penetrate deeper into the regions of each other, resulting
in
a bigger additional widening of the reaction zone}. This slowing down of the
annihilation process is probably due to the inability of diffusion to destroy
fluctuations beyond a certain length scale, as in the homogeneous case above.
The analogy cannot be upheld, however, beyond the length scale set by the size
of the depletion zone, which is absent in the homogeneous system.

\subsection{Complex reaction-diffusion systems}
\indent

We now consider reaction-diffusion systems which are significantly more complex
than the ones considered so far. They involve species with different transport
properties (e.g.\ diffusion coefficients) and various chemical reactions, some
of which are autocatalytic, i.e.\ they require the presence of a certain
species in order to produce more of it. Autocatalytic reactions play an
important r\^ole in biological processes. Systems subject to autocatalytic
reactions can undergo phase transitions far from thermodynamic equilibrium.
A system can be kept far from chemical equilibrium, for example, by suppressing
reverse reactions and/or by an external supply of reactants. In such systems,
the (unique) steady state, which is stable near equilibrium, may become
unstable
as certain parameters are varied. Then a phase transition to a new state may
take place. Thus, an originally homogeneous state may become unstable and
spatial concentration patterns (Turing structures) may develop spontaneously
(see e.g.~\cite{np,m}). A crucial element in the formation of Turing structures
is a significant difference in the diffusion coefficients of two species; the
faster species, the {\em inhibitor}, hinders by chemical action the spreading
of
the slower one, the {\em activator}, and the latter accumulates in restricted
areas, creating a pattern of inhomogeneous concentration. The experimental
observation of a sustained standing non-equilibrium chemical pattern has been
reported recently~\cite{cdbk}. The possibility of such distinct qualitative
behaviour makes autocatalytic systems attractive as a testing ground of the
model developed here. It should be clear at the outset that the formation of
structured states is predicted by the macroscopic diffusion-reaction equations.
Linear stability analysis provides, in fact, a critical value, $d_c$, of the
ratio of the inhibitor diffusion coefficient to that of the activator; when
the ratio increases beyond $d_c$, a range of Fourier components of the
concentration become unstable and appear as spatial oscillations with
the corresponding wavelengths. The details of the transition, such as the value
of $d_c$, may depend, however, on microscopic fluctuations and differences
between the differential equation and CA approaches may appear~\cite{df}. Our
aim here is to show that our model is universal enough to describe qualitative
phenomena of the kind described above.

As a first example we consider the Brusselator~\cite{pl}, defined by the
reaction scheme
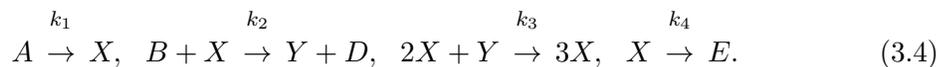
\begin{equation}
A \begin{array}[b]{c} \mbox{\scriptsize $k_1$} \\ \rightarrow \end{array} X ,
\hspace{.25cm}
B + X \begin{array}[b]{c} \mbox{\scriptsize $k_2$} \\ \rightarrow \end{array}
Y + D, \hspace{.25cm}
2X + Y \begin{array}[b]{c} \mbox{\scriptsize $k_3$} \\ \rightarrow \end{array}
3X, \hspace{.25cm}
X \begin{array}[b]{c} \mbox{\scriptsize $k_4$} \\ \rightarrow \end{array} E.
\end{equation}
Here the concentrations of $A$ and $B$ are kept constant and $D$ and $E$ are
continuously removed. In this case, $X$ is the activator and $Y$ the inhibitor.
The Brusselator, albeit relatively simple, displays the striking qualitative
behaviour (concentration oscillations in time and space, nonlinear travelling
waves) exemplified by the Belousov-Zhabotinski reaction~\cite{bz}. We consider
a one-dimensional system  of length $L$. Following Ref.~\cite{m}, one can put
the reaction-diffusion equations for the unconstrained concentrations in a
dimensionless form,
\begin{eqnarray}
\addtocounter{zaehler}{1}
{\partial \tilde{C}_X \over \partial \tilde{t}} & = &
\tilde{D}_X {\partial^2 \tilde{C}_X \over \partial \tilde{x}^2} +
\tilde{C}_A - (\tilde{C}_B + 1) \tilde{C}_X + \tilde{C}_X^2 \tilde{C}_Y \\
\addtocounter{equation}{-1} \addtocounter{zaehler}{1}
{\partial \tilde{C}_Y \over \partial \tilde{t}} & = &
\tilde{D}_Y {\partial^2 \tilde{C}_Y \over \partial \tilde{x}^2} +
\tilde{C}_B \tilde{C}_X - \tilde{C}_X^2 \tilde{C}_Y \hspace{0.25cm} ,
\end{eqnarray}
\setcounter{zaehler}{0}
by the transformation $\tilde{t} \equiv k_4 t$, $\tilde{x} \equiv x/L$,
$\tilde{D}_{X,Y} \equiv D_{X,Y}/k_4 L^2$, $\tilde{C}_A \equiv (k_1 k_3^{1/2} /
k_4^{3/2}) C_A$, $\tilde{C}_B \equiv (k_2 / k_4) C_B$ and $\tilde{C}_{X,Y}
\equiv (k_3 / k_4)^{1/2} C_{X,Y}$. Since $C_B$ is held fixed and $D$
plays no active r\^ole, the second reaction in the original scheme can be
effectively replaced by $X \rightarrow Y$, with rate constant $k_2 C_B$.

\begin{figure}[tp]
  \vspace{9.0cm}
  \caption{Simulation of one-dimensional Brusselator (solid curve) compared
with
  solution of reaction-transport equations (dotted curve).
  \label{fig8} }
  \vspace{9.0cm}
  \caption{Spatial concentration pattern obtained from simulation of
  two-dimensional Schnackenberg model; regions of high (low) density are
  indicated in red (blue).
  \label{fig9} }
\end{figure}

Fig.~\ref{fig8} shows the result of a simulation of
Eqs.~(\thesection.\arabic{equation}) on a one-dimensional lattice of length
$1.01 L$ (1010 sites, $\tilde{\lambda} \equiv \lambda / L = 10^{-3}$). We
choose
the physicochemical parameters to have the values quoted in Fig.~7.13 of
Ref.~\cite{np}: $\tilde{C}_A = 2$, $\tilde{C}_B = 4.6$, $\tilde{D}_X = 1.6
\times 10^{-3}$ and $\tilde{D}_Y = 3.75 \, \tilde{D}_X$ ($d_c = 3.05$). The
dimensionless concentrations $\tilde{C}_{X,Y}$ are computed by averaging the
particle densities over cells of 10 sites and multiplying them by a scaling
factor $\gamma \simeq 0.2$; to obtain $C_{X,Y}$, we multiply by an additional
factor $(k_4/k_3)^{1/2}$. We constrain the boundary concentrations
$\tilde{C}_X$ and $\tilde{C}_Y$ to remain fixed at the values of the (unstable)
homogeneous steady state, i.e.\ $\tilde{C}_A$ and $\tilde{C}_B / \tilde{C}_A$
respectively; this is done as follows: before every iteration step, we replace
all particles in the first and last cells by uniform random distributions of
particles having the necessary density. Initially $X$- and $Y$-particles are
distributed uniformly in the interior cell region of the lattice with densities
of 9.2 and 10.7 particles/site respectively. The above value of $\gamma$ is
chosen to give initial concentrations of 1.8 and 2.1 respectively. In the
simulation we face a problem of competing reactions (e.g.\ a single
$X$-particle
may either convert to a $Y$-particle or decay). Thus, in a naive
implementation,
in which all reactions are given a chance to take place one after the other,
the
order in which this is done would be significant. We solve the problem by
dividing up reactions in groups of competing reactions and allowing randomly
only one reaction from each group to take place during an iteration step (the
probability of each reaction has to be multiplied of course by the number of
members in its group). For the  parameter values chosen, the initial
homogeneous
state is unstable and the system evolves to a structured state. The spatial
structure shown is obtained after $800 \, 000$ iterations with time step
$\tilde{\tau} \equiv k_4 \tau =  0.833 \times 10^{-4}$. Obviously the
statistics
is still poor for a quantitative comparison with the continuum result, which,
incidentally, has converged to a stationary value by the time shown here.

The concentration pattern shown in Fig.~\ref{fig9} is produced with the
reaction
scheme
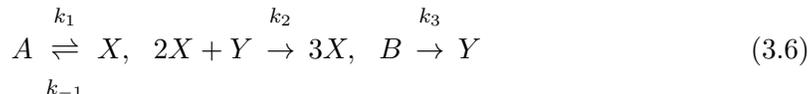
\begin{equation}
A \begin{array}{c} \mbox{\scriptsize $k_1$} \\ \rightleftharpoons \\
\mbox{\scriptsize $k_{-1}$} \end{array} X, \hspace{.25cm}
2X + Y \begin{array}[b]{c} \mbox{\scriptsize $k_2$} \\ \rightarrow \end{array}
3X, \hspace{.25cm}
B \begin{array}[b]{c} \mbox{\scriptsize $k_3$} \\ \rightarrow \end{array} Y
\label{schnack}
\end{equation}
(Schnackenberg model~\cite{s}). The simulation is performed on a $696 \times
721$ lattice of size $1.10 \times 1.14$ (arbitrary units). Initially $X$- and
$Y$-particles are distributed randomly, with uniform density 1.55 and 0.59
particles/site respectively. Other parameters are chosen as in Fig.~2a of
Ref.~\cite{bd}: $C_B = 1.41 \, k_1^{3/2} / k_3 k_2^{1/2}$, $C_A = 0.14 \,
k_1^{3/2} / k_{-1} k_2^{1/2}$, $D_X = 10^{-4} \, k_1 L^2$ ($L$ being the linear
size of the system). Finally we take $D_Y = 30 \, D_X$ ($d_c \simeq 20$). The
density of $X$-particles is shown here after $4 \, 800$ iterations, with $\tau
=
0.625 \times 10^{-6} \, k_1^{-1}$. Time limitations have prevented us from
running the simulation up to a time when the pattern becomes stationary.
CA simulation of spatial patterns with the Schnackenberg model has recently
been
reported by another group~\cite{df}. \footnote{We have also performed
simulations of the Selkov reaction scheme, which was introduced in the context
of glycolytic oscillations and is obtained from (\ref{schnack}) by making all
reactions reversible~\cite{se}. We obtain concentration patterns, in agreement
with Ref.~\cite{klm}, both above and below the critical diffusion-coefficient
ratio. Below the critical ratio, the homogeneous state is apparently
destabilised by density fluctuations which effectively widen the range of
unstable wavenumbers.}

\newpage
\section{Conclusion and Outlook} \label{C-O}
\indent

Reaction-transport processes were modelled in this work as a cellular
automaton.
Particles are transported by executing a random walk on the sites of a regular
lattice and are chemically transformed according to a local probabilistic rule.
The microscopic random motion of the particles is manifested, at the
macroscopic
level, as a combination of advection and diffusion. In particular, advection
arises from a directional bias in the random walk, i.e. if particles have a
relatively higher probability to move in the direction of the advection
velocity. Chemical reactions are likewise modelled at the microscopic level: in
the process of the random walk, we allow particles that meet at a lattice site
the chance of `reacting', i.e.\ disappearing and leaving in their wake a set of
new particles, the products of the reaction. The model is {\em simple} and {\em
general}. The evolution equations are transcribed into simple computer code.
This simplicity does not preclude, but, on the contrary, facilitates the
implementation of arbitrarily complex reactions and boundary conditions, in a
physically transparent way. The results of the previous section demonstrate
that
the model presented here {\em can successfully describe a wide variety of
reaction-transport systems}.

In the continuum limit, the evolution equations of the discrete model go over
to
the standard reaction-transport PDE's, if certain conditions are fulfilled,
namely if molecules of different species are uncorrelated (molecular chaos) and
if particle density is smooth in space. These conditions are, however, not
imposed in our model; thus our simulations  account for microscopic effects
(e.g.\ fluctuations) that are typically averaged out by the continuum approach.
The results of the discrete model were compared carefully to the solution of
the
reaction-transport PDE's in the case of a simple homogeneous system of
particles. We encountered two sources of discrepancy: on the one hand
correlations between the particles, that have no physical significance and are
mere artifacts of the discretisation, and, on the other hand, statistical
fluctuations that influence the long-time behaviour of reaction-diffusion
systems. We emphasise the latter kind of discrepancy, which is of a fundamental
nature and shows that microscopic fluctuations can influence qualitatively the
evolution of macroscopic systems. The differences between the discrete
simulation and the continuum approach in the time development of the reaction
front forming when the reacting species are initially separated is probably of
similar origin. Macroscopic consequences of microscopic fluctuations were also
seen in the case of autocatalytic reaction-diffusion systems, near the
threshold
for the onset of non-equilibrium phase transitions leading to formation of
spatially structured states. We have thus shown in concrete cases that the
discrete model {\em can be used to approximate systematically the respective
PDE's, while, unlike the latter, it accounts for effects of explicitly
microscopic origin}.

An important aspect of our approach is that the same macroscopic behaviour can
be
obtained with a variety of microscopic rules. This freedom is inherent in CA
modelling, since the aim is to model only certain fundamental features of the
microscopic world, but not the full  detail of the dynamics. As examples, we
gave two such rules: rule I, a simple rule leading to the standard rate law
only in the limit of small densities, and rule II, a more complicated rule
that, however, results in the standard rate law for any density. The
computation
time required by different rules in order to simulate the same macroscopic
behaviour may vary. Our comparison of  chemical rules I and II indicates that
the latter affords better possibilities  of optimisation, due to the wider
range
of possible particle densities.

Finding the optimal microscopic rule is one way of enhancing computational
efficiency, which is an essential task in practical applications. As a
consequence of the unlimited number of particles per site, our model does not
vectorise well when implemented on computers with vector architecture. In the
present general form of the model, the best scalar performance is attained on a
VAX-9000 and amounts to $\sim 1 \, 000 \, 000$ site updates per second (u.p.s.)
for pure transport of one species, $\sim 100 \, 000$ u.p.s.\ for $a + b
\rightleftharpoons c$ and $\sim 20 \, 000$ u.p.s.\ for the Brusselator. This
performance can be improved by varying the particle density, as we saw in
Section~\ref{simul}. However, our model should be implemented on a massively
parallel computer in order to achieve its full potential. The great advantage
of
our model is its ability to model a very wide range of reaction-transport
problems. The performance figures quoted above are thus for a correspondingly
general code. For specific applications, both the model and corresponding code
can be appropriately tailored to optimise performance still further (see for
example Ref.~\cite{cd} where $20 \, 000 \, 000$ u.p.s.\ were achieved on a
CRAY-YMP for a specific reaction-diffusion process).

An important feature of our CA approach is its inherent stability. Indeed our
algorithm is stable even for values of $\tau$ and $\lambda$ for which the
corresponding FDE's, represented by Eqs.~(\ref{FD-full}) and
(\ref{evol-tran1}),
are numerically unstable when solved by direct iteration with particle
densities
being floating point variables. We note that adding chemical reactions to a
problem of pure transport can turn a stable FDE algorithm into an unstable one
- for the same time and space discretisation \footnote{Stability is then
achieved by refining further the time discretisation.}. In such cases, where
deterministic methods using floating point numbers fail, our stochastic
approach, using integer variables, provides a stable method of solving
Eqs.~(\ref{FD-full}) and (\ref{evol-tran1}). The lack of general stability
criteria for FDE in the presence of chemical reactions makes this guaranteed
stability of our approach a valuable asset.

Realistic applications will constitute the principal direction of further work
on our model. These applications will be chosen on grounds of practical
usefulness and so as to utilise the advantages of our approach: the guaranteed
numerical stability and the capacity to treat arbitrary boundary conditions and
chemical systems which are not necessarily in chemical equilibrium. Thus
applications will include both problems at field scale, where differential
equations are used widely to describe average quantities, as well at the scale
of the pores, which form networks with very complex boundary conditions. As
examples we mention the chemistry of mineral surfaces
(precipitation/dissolution) and the effects of porosity changes on mass
transport \cite{appl}.

In conclusion, we have shown that the cellular automaton model we are proposing
is capable of simulating simple as well as complex reaction-transport
processes.
We understand well its relation to other numerical methods for solving the
corresponding differential equations and will proceed to apply it to realistic
problems. Where other methods work as well, the relevant question will be that
of the relative computational efficiency. More challenging will be, however, to
identify areas of application where differential equation approaches are
confronted with either serious numerical or even conceptual problems.

\subsection*{Acknowledgements}
\indent

We would like to thank J. Hadermann for his interest in our work and C. L.
Carnahan for a critical reading of the manuscript. Partial financial support by
NAGRA is gratefully acknowledged. One of the authors (T.K.) has benefited a
great deal from discussions with B. Chopard, M. Droz, L. Frachebourg, J.-P.
Boon
and D. Dab.

\end{document}